\documentclass[journal]{IEEEtran}
\usepackage{flushend}
\usepackage{balance}

\usepackage{booktabs}

\usepackage[colorlinks = true,
            linkcolor = black,
            citecolor = black,
            urlcolor = blue]{hyperref}

\usepackage{cite}

\usepackage{color}

\ifCLASSINFOpdf
	\usepackage[pdftex]{graphicx}
\else
\fi
\usepackage{amsmath}
\usepackage{amsfonts}

\ifCLASSOPTIONcaptionsoff
 \usepackage[nomarkers]{endfloat}
\let\MYoriglatexcaption\caption
\renewcommand{\caption}[2][\relax]{\MYoriglatexcaption[#2]{#2}}
\fi
\newcommand{\reb}[1]{\textcolor{black}{#1}}
\newcommand{\minorrev}[1]{\textcolor{black}{#1}}
\newcommand{\changes}[1]{\textcolor{black}{#1}}

\begin{document}
%
\title{DeepISP: \minorrev{Towards} Learning an End-to-End Image~Processing~Pipeline}
%
%
%
\author{Eli Schwartz, Raja Giryes and Alex M. Bronstein
\thanks{E. Schwartz and R. Giryes are with the School of Electrical Engineering, Tel-Aviv University, Israel. \{eliyahus@mail,raja@tauex\}.tau.ac.il. A. Bronstein is with the Department of Computer Science, Technion -- Israel Institute of Technology, Israel. bron@cs.technion.ac.il.}
}
\maketitle

\begin{abstract}%
We present DeepISP, a full end-to-end deep neural model of the camera image signal processing (ISP) pipeline. Our model learns a mapping from the raw low-light mosaiced image to the final visually compelling image and encompasses low-level tasks such as demosaicing and denoising as well as higher-level tasks such as color correction and image adjustment. The training and evaluation of the pipeline were performed on a dedicated dataset containing pairs of low-light and well-lit images captured by a Samsung S7 smartphone camera in both raw and processed JPEG formats. The proposed solution achieves state-of-the-art performance in objective evaluation of PSNR on the subtask of joint denoising and demosaicing. For the full end-to-end pipeline, it achieves better visual quality compared to the manufacturer ISP, in both a subjective human assessment and when rated by a deep model trained for assessing image quality.%
\end{abstract}


%
\IEEEpeerreviewmaketitle

\begin{figure*}[htb]   
    \centering  
    \begin{tabular}{c@{\hskip 0.01\textwidth}c@{\hskip 0.01\textwidth}c@{\hskip 0.01\textwidth}c@{\hskip 0.01\textwidth}c@{\hskip 0.01\textwidth}c}
        \multicolumn{2}{c}{\includegraphics[width = 0.3\textwidth]{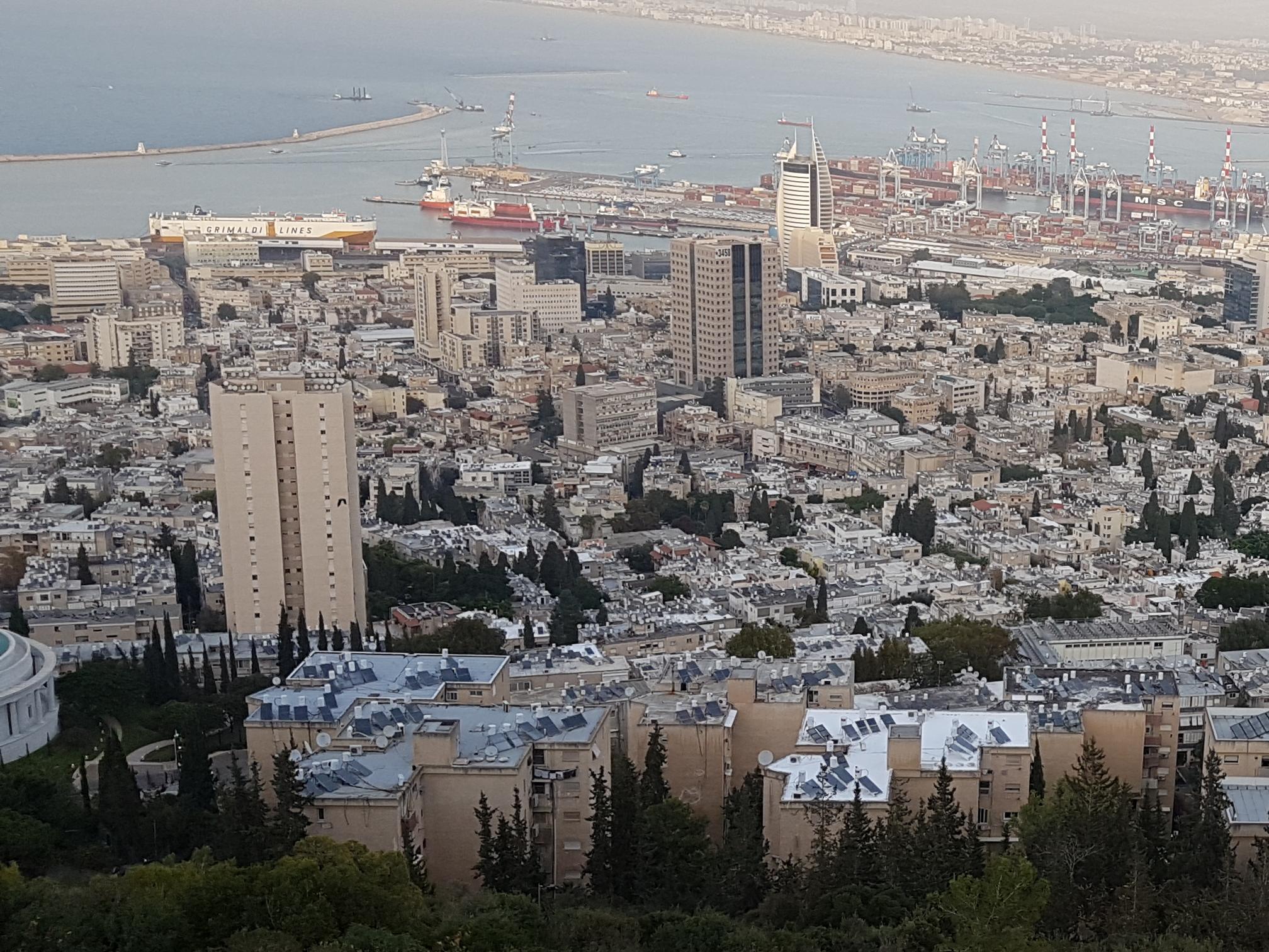}} &
        \multicolumn{2}{@{}c}{\includegraphics[width = 0.3\textwidth]{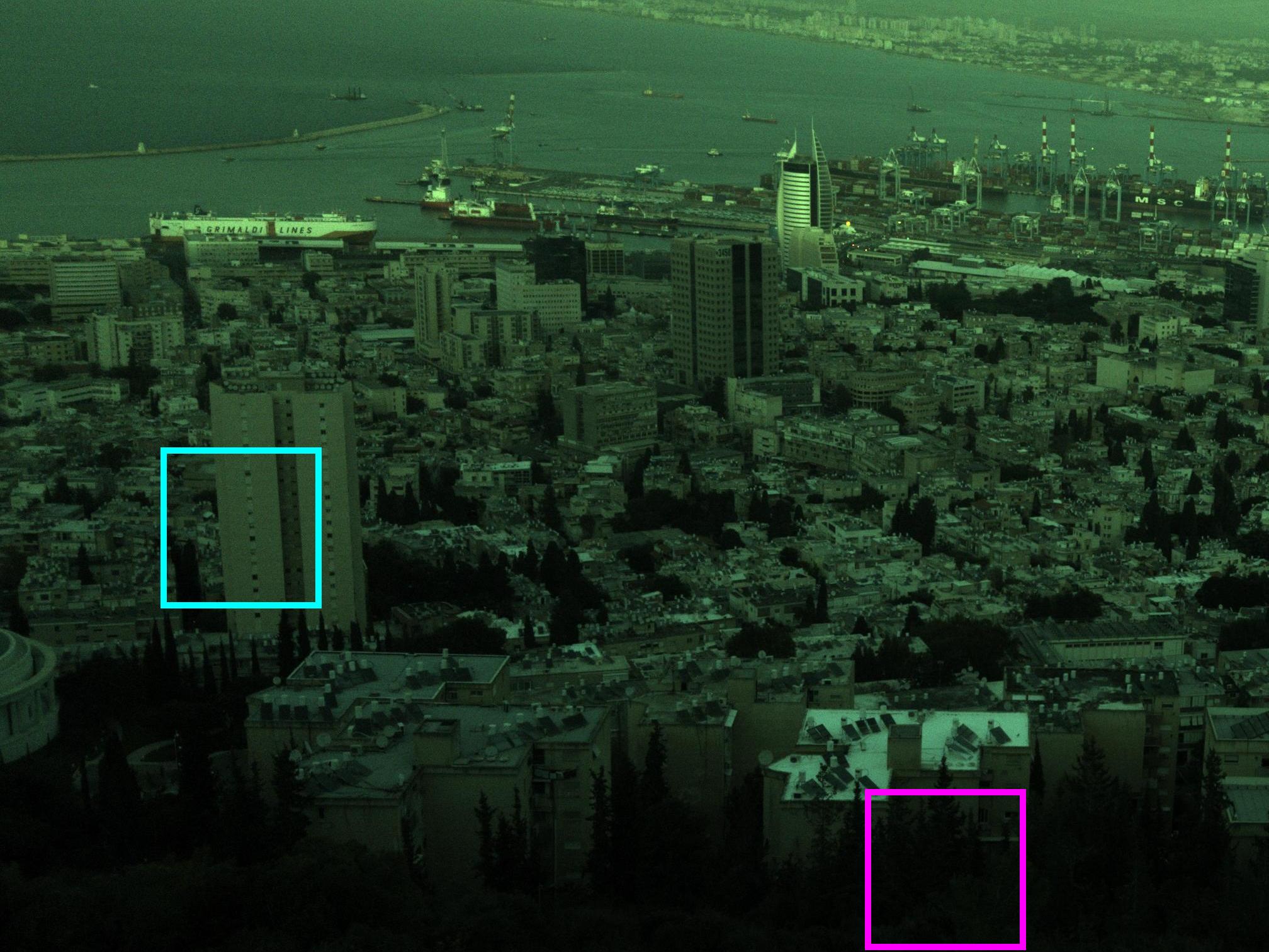}} &
        \multicolumn{2}{@{}c}{\includegraphics[width = 0.3\textwidth]{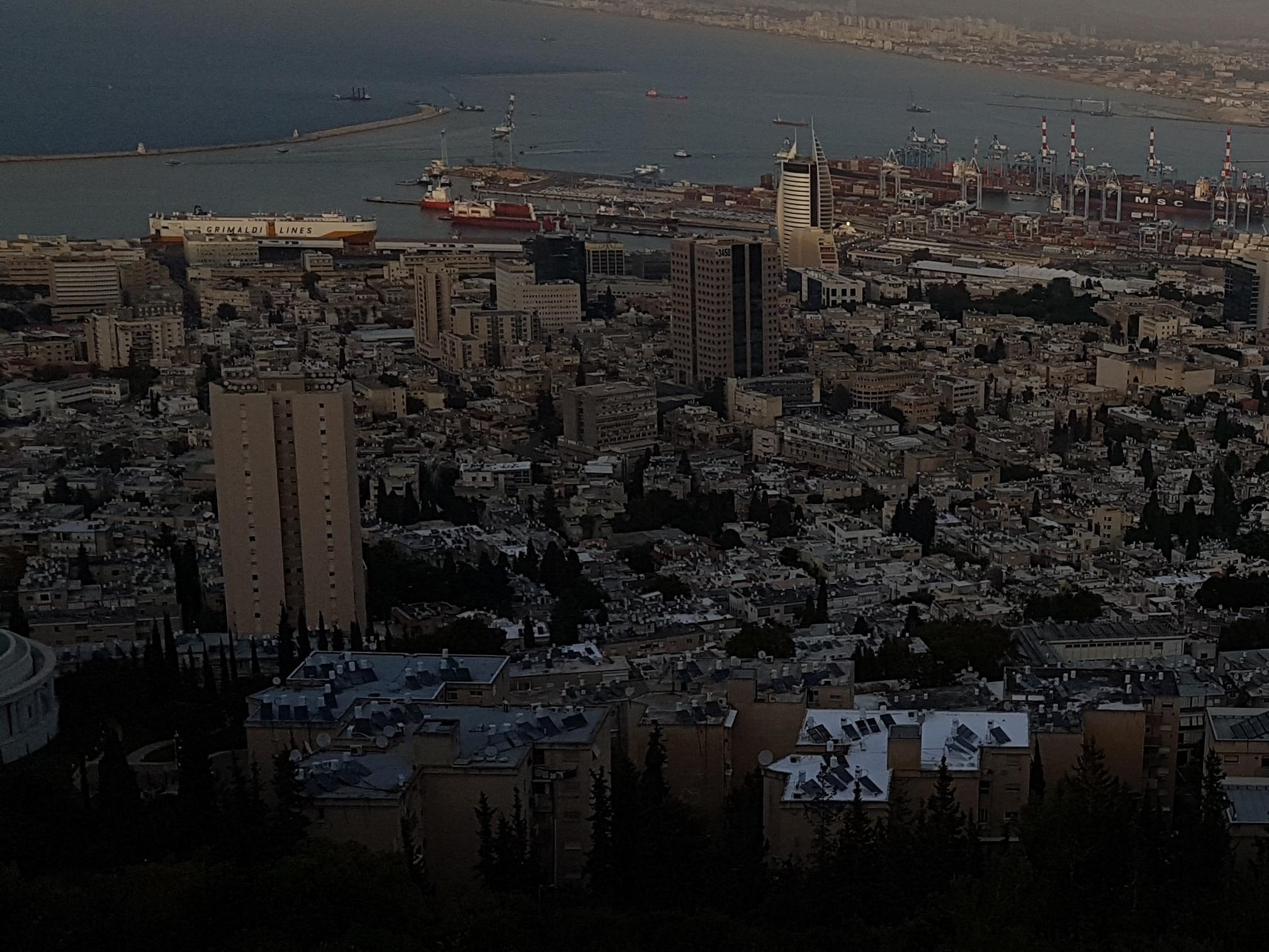}} \\
        
    \includegraphics[width = 0.145\textwidth]{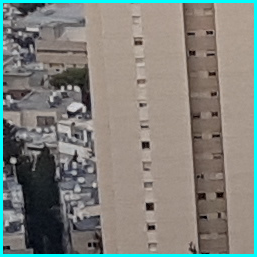} &
    \includegraphics[width = 0.145\textwidth]{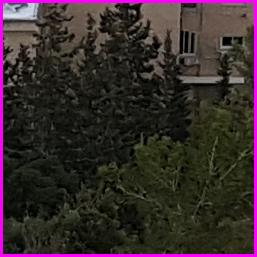} &
    \includegraphics[width = 0.145\textwidth]{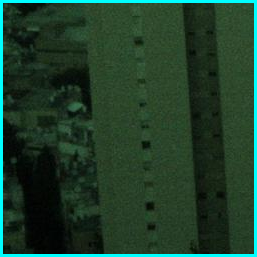} &
    \includegraphics[width = 0.145\textwidth]{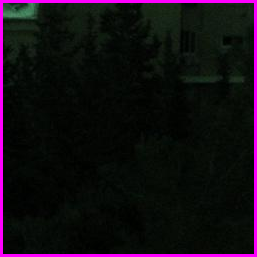} &
    \includegraphics[width = 0.145\textwidth]{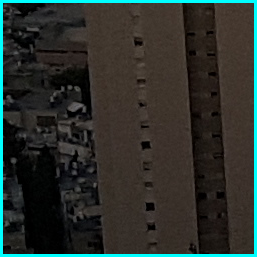} &
    \includegraphics[width = 0.145\textwidth]{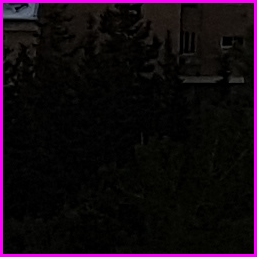} \\
    
    \multicolumn{2}{c}{Well-lit -- Used as ground truth} &
    \multicolumn{2}{@{}c}{Low-light raw input} &
    \multicolumn{2}{@{}c}{Samsung S7} \\
    \multicolumn{2}{c}{MOS=3.7} &
    \multicolumn{2}{@{}c}{} &
    \multicolumn{2}{@{}c}{MOS=2.2} \\
    
            {} & \multicolumn{2}{@{}c}{\includegraphics[width = 0.3\textwidth]{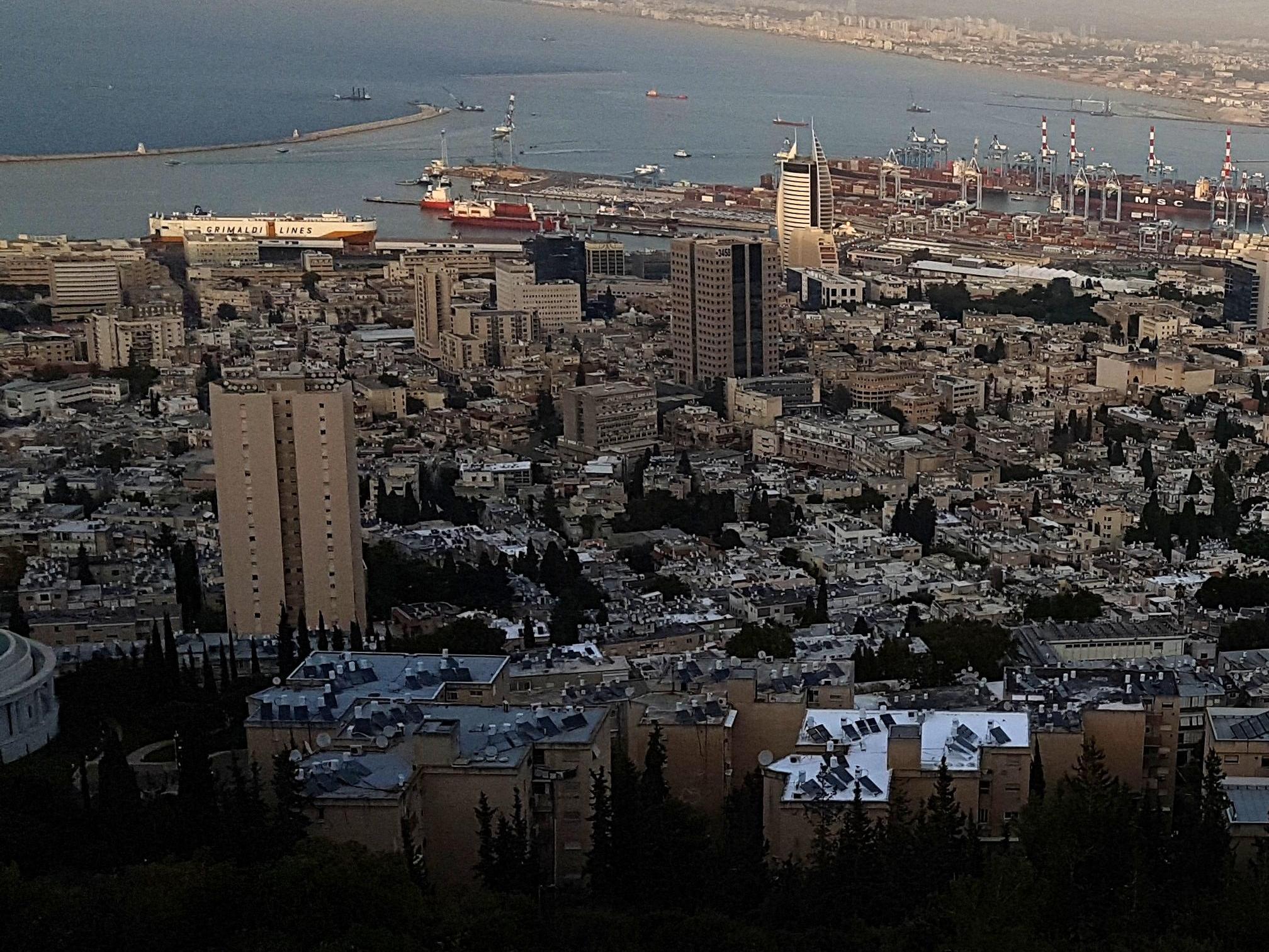}} &
        \multicolumn{2}{@{}c}{\includegraphics[width = 0.3\textwidth]{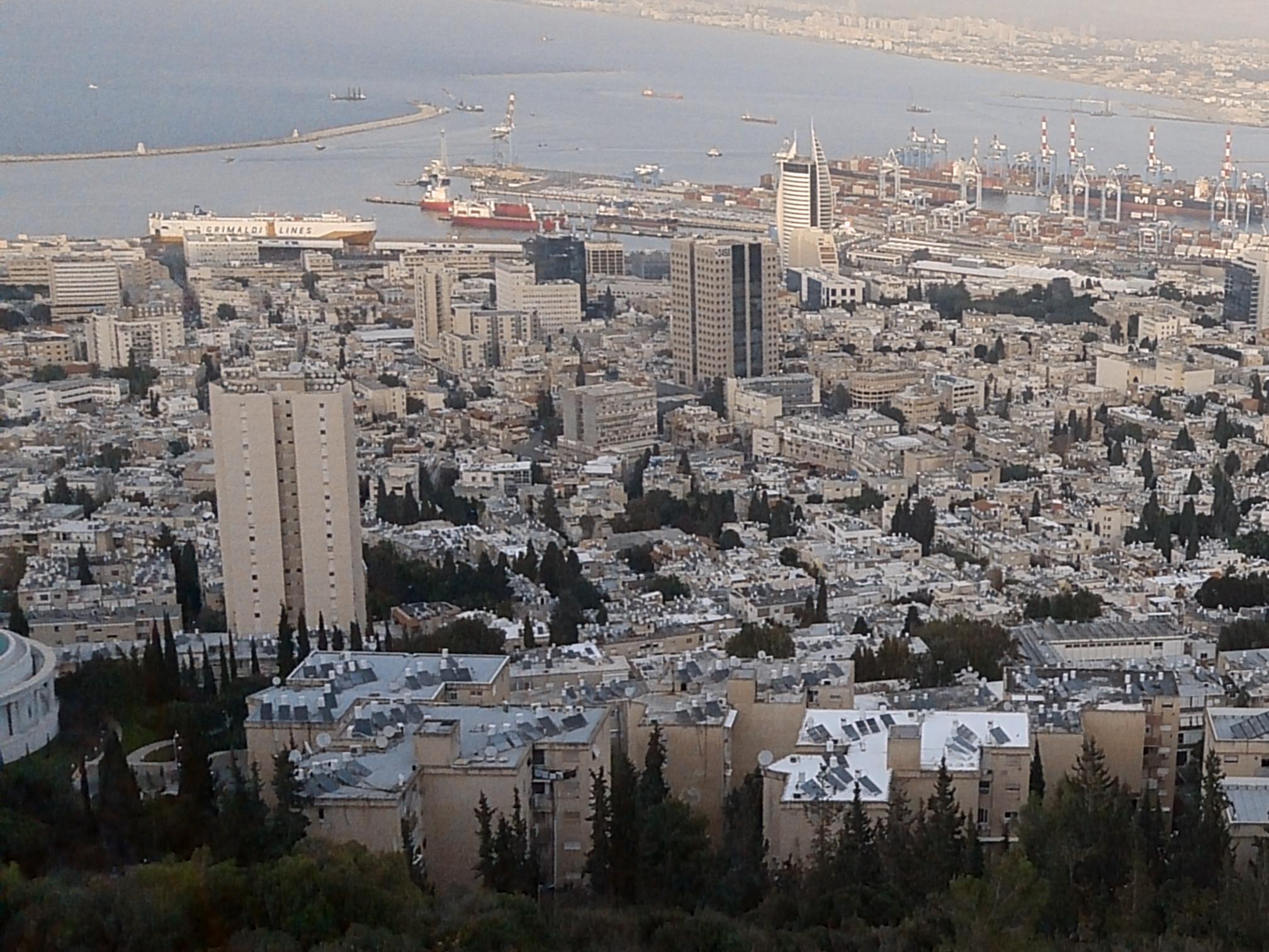}} & {} \\
        
    {} &
    \includegraphics[width = 0.145\textwidth]{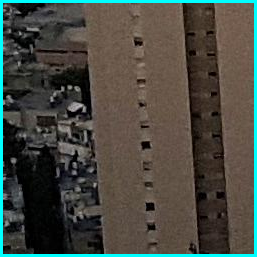} &
    \includegraphics[width = 0.145\textwidth]{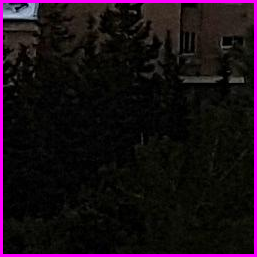} &
    \includegraphics[width = 0.145\textwidth]{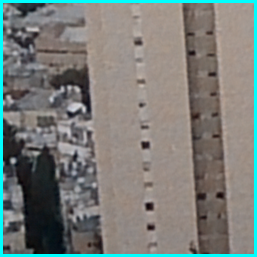} &
    \includegraphics[width = 0.145\textwidth]{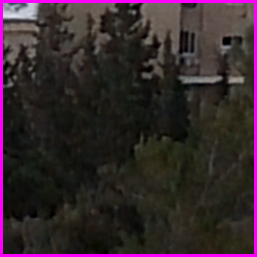} &
    {} \\
    
    {} &
    \multicolumn{2}{@{}c}{Samsung S7 (Histogram stretched)} &
    \multicolumn{2}{@{}c}{DeepISP} &
    {} \\
    {} &
    \multicolumn{2}{@{}c}{MOS=3.3} &
    \multicolumn{2}{@{}c}{MOS=3.6} &
    {} \\
    \end{tabular} 
   
    \smallskip 
    \caption{\small \textbf{End-to-end low-light image processing.} \minorrev{Top-down left-right: well-lit image used as our training ground truth, raw input low-light image (for visualization purposes, after demosaicing by bilinear interpolation), output of the Samsung S7 ISP, output of the Samsung S7 ISP after histogram stretch to re-map to the full 0-255 range (used for visualization and evaluation), and of the proposed DeepISP (the same histogram stretch is applied). Mean Opinion Score (MOS) is displayed below each image.} More examples are provided in Fig. \ref{fig_teaser_extra}.}
    \label{fig_teaser}
\end{figure*}

\section{Introduction}

\IEEEPARstart{I}{n} recent years, the use of high resolution cameras in mobile phones has become increasingly popular. However, due to space constraints, their hardware is limited with respect to the pixel size and the quality of the optics. Moreover, mobile phones are usually hand-held, thus, not stable enough for long exposure times. Therefore, in these devices the imaging hardware must be paired with capable algorithms to compensate for these limitations. For these reasons, the focus of this work is on low-light mobile image processing. In particular, our goal is to propose a radical alternative to the existing ISP (image signal processor) in such devices. 

ISP is a special hardware in cameras dedicated to image processing tasks. It is also used as a synonym for the image processing pipeline. 
This pipeline encompasses a sequence of operations, ranging from \emph{low-level} demosaicing, denoising and sharpening to \emph{high-level} image adjustment and color correction.
Typically, each task is performed independently according to different heavily engineered algorithms per task.

Deep learning (DL)-based methods, and more specifically convolutional neural networks (CNNs), have demonstrated considerable success in such image processing tasks. For example, these models have produced state-of-the-art results for demosaicing \cite{gharbi2016deep,klatzer2016learning}, denoising \cite{remez2017deep, zhang2017learning, burger2012image,Feng15Fast,Chen16Trainable,vemulapalli2016deep, Zhang17Beyond}, deblurring \cite{Schuler16Learning, Xu14Deep, Sun15Learning, Chakrabarti16Neural, su2016deep, Nah17Deep} and super-resolution \cite{Kim16Accurate, Bruna16Super, ledig2016photo, tong2017image, Lim17Enhanced}. Traditional image processing algorithms commonly rely on hand-crafted heuristics, which require explicitly defining the prior on natural images statistics. Some examples of priors used in the literature are: a sparse representation in a redundant dictionary \cite{aharon2006rm}, local smoothness \cite{rudin1992nonlinear} and non-local similarity \cite{buades2005non}. An advantage of DL-based methods is their ability to implicitly learn the statistics of natural images. Moreover, recent research has demonstrated that CNNs are inherently good at generating high-quality images, even when operating outside the supervised learning regime, e.g.,  \cite{bahat2017non} and \cite{ulyanov2017deep}.

Some studies have explored the application of deep learning for other image enhancement tasks. For example, the mapping between pairs of dark and bright JPEG images is learned in \cite{shen2017msr}. Another example is learning a mapping from mobile camera images to DSLR images \cite{ignatov2017dslr}. Note that these two works (among others) do not perform end-to-end processing. Instead, they begin from an image already processed by an ISP.

Conditional generative adversarial networks (CGANs) are another common approach for image enhancement. These models consist of a generator and a discriminator. The generator maps the source domain distribution to an output domain distribution given an input image. The learning is accomplished by having a discriminator that learns to distinguish between generated images and real images and optimizing the generator to fool the discriminator. In one example, color correction for underwater images was learned using such an adversarial loss (in addition to other loss terms) \cite{li2017emerging}. More examples of GAN used for image restoration are super-resolution \cite{ledig2016photo} and blind deblurring \cite{Nah17Deep}. A main limitation of GANs is that they are not very stable in training and tend to suffer from mode collapse, so only a subset of the domain distribution is generated. For this reason, the use of GANs for image enhancement requires adding other loss terms.

In contrast to the traditional approach that solves independently the sequence of tasks performed in the standard ISP, DL allows to jointly solve multiple tasks, with great potential to alleviate the total computational burden. Current algorithms were only able to accomplish this for closely-related tasks, such as denoising and demosaicing \cite{khashabi2014joint} or super-resolution and demosaicing \cite{farsiu2006multiframe}. These studies have shown the advantage of jointly solving different tasks. In this paper, we demonstrate the ability to jointly learn in an end-to-end fashion the full image processing pipeline. Such an approach enables sharing information (features) between parts of the network that perform different tasks, which improves the overall performance compared to solving each problem independently.

\textbf{Contribution.}
This paper has three main contributions: 
\begin{itemize}
\item Firstly, a novel deep neural model is proposed for low-level image enhancement that achieves state-of-the-art results for joint denoising and demosaicing. 
\item Secondly, we extend the model to an end-to-end image processing pipeline that we name DeepISP, which accepts a raw image as the input and outputs a final high perceptual quality image.
\item Finally, we release to the public domain the S7 ISP dataset, containing raw and processed JPEG images of the same scenes captured by a Samsung S7 phone camera with normal (automatically chosen) exposure time and shorter exposure time simulating low-light conditions.
\end{itemize}

\section{The DeepISP network}
We turn to describe now our proposed data-driven solution for the image processing pipeline. The model jointly learns low-level corrections, such as demosaicing and denoising, and higher level global image restoration in an end-to-end fashion. Our motivation is that different tasks in this pipeline can be performed better when performed simultaneously. In addition, it has an advantage with respect to computational efficiency as computed features are shared between the different tasks. 

\subsection{Network Architecture}

\begin{figure}[!t]
\begin{center}
   \includegraphics[width=0.8\linewidth]{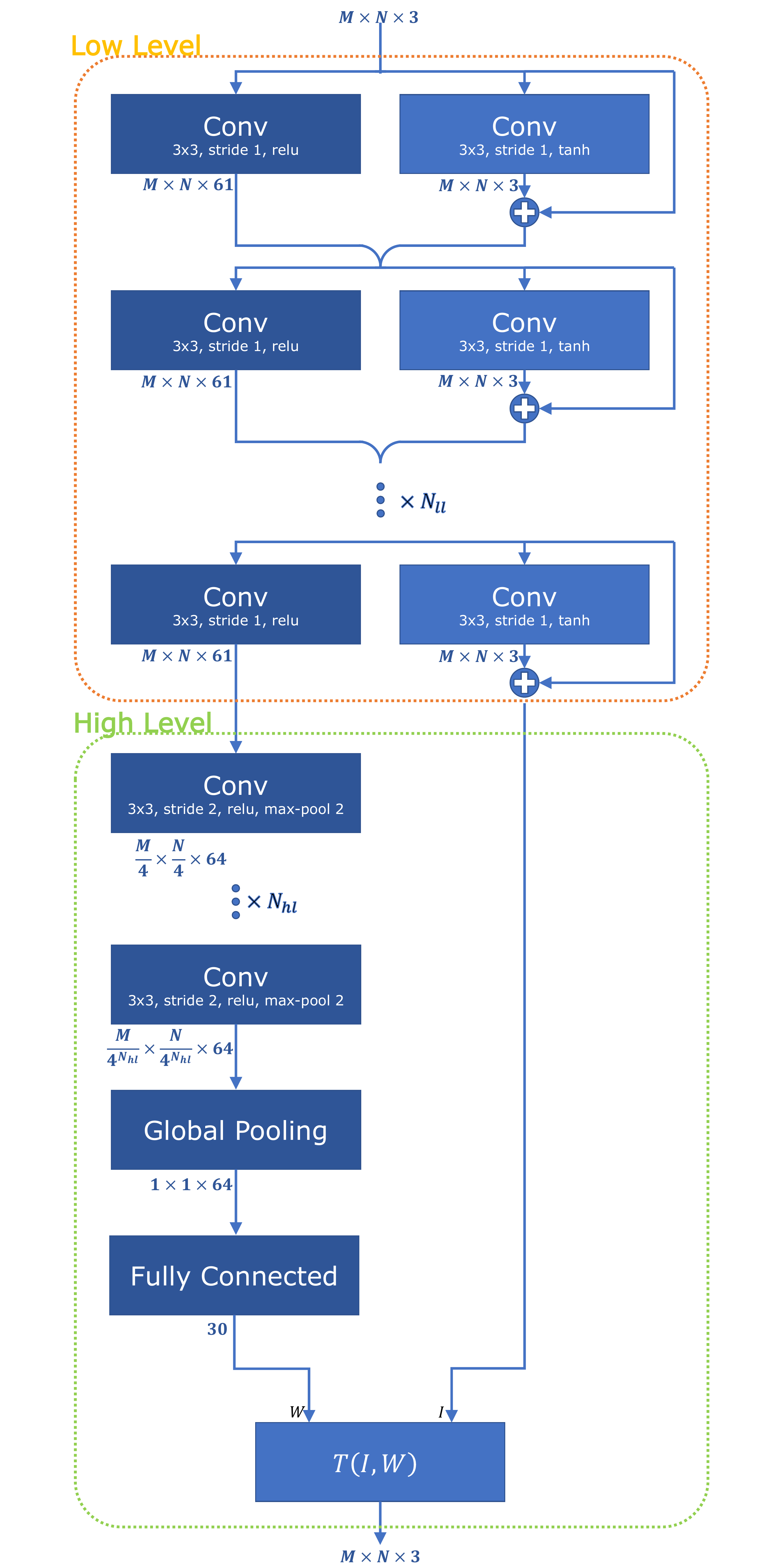}
\end{center}
       \caption{\textbf{Proposed network architecture.} The network consists of two stages: Lower level and higher level. Layers that output features are colored dark blue and layers that output an image (or residual image) are colored bright blue.}
       \label{fig_net_architecture}
    \label{fig:long}
    \label{fig:onecol}
\end{figure}

Fig.~\ref{fig_net_architecture}
presents our proposed network architecture for an end-to-end image processing and enhancement, denoted as DeepISP.
DeepISP is composed of two stages, depicted in orange and green bordures in the diagram, respectively. The first stage extracts low-level features and performs local modifications. The second one extracts higher level features and performs a global correction. The network is fully convolutional, thus, can accommodate any input image resolution.

\paragraph{Low-level stage} The low-level part of DeepISP consists of $N_{ll}$ blocks.
Each intermediate block performs convolution with filters of size $3 \times 3$  and stride $1$. Its input and output sizes are $M \times N \times 64$, where $M$ and $N$ are the input image dimensions. The same dimensions are maintained by applying reflection padding to the input. The input to the network is a demosaiced RGB image produced by a simple bilinear interpolation in a preprocessing stage.

At each layer, $61$ out of the $64$ channels are standard feed-forward features (left column in the diagram). The other $3$ channels contain a correction for the RGB values of the previous block, i.e., they contain a residual image that is added to the estimation of the previous layer. 
Clearly, the first block of the network has a similar structure but with only the $3$ channels of the input image (and not $64$ as the other blocks).
\reb{
ReLU nonlinearity is used for the feature blocks, as common in recent DL models. However, for the residual images we would like to be able to predict both positive and negative values and thus we use the $\tanh$ nonlinearity.
}

The architecture of the low-level part is inspired by the one suggested in \cite{remez2017deep} for denoising. Similarly to \cite{remez2017deep}, small convolutions are applied and each block produces a residual image. Unlike \cite{remez2017deep}, where all residual images are accumulated at the last layer, we perform the summation at each block, thus, allowing the network to get at each level the current image estimate in addition to the calculated features.

\paragraph{High-level stage} 

The last block at the low-level stage forwards the $61$ feature channels in one path and the currently estimated image ($I$) in another path to the high-level stage of the network.
The latter uses the features from the low-level stage for estimating a transformation $W$ that is then applied to the image corrected by the first stage ($I$) to produce a global correction of the image. 

This stage includes a sequence of $N_{hl}$ convolution layers with filters of size $3 \times 3$  and stride $2$. Each layer is followed by a $2 \times 2$ max-pooling. The purpose of the strides and pooling is getting a large receptive field and lowering the computational cost. 
A global mean-pooling is applied to the output of these convolutions, resulting in a single feature vector. This is followed by a fully connected layer that produces the parameters of the transformation $W$.  

In this work, we use a quadratic function of the pixel's R, G, and B components; it is applied pixel-wise as
\begin{eqnarray}
W \cdot triu \left( \left[ \begin{array}{cccc} r &  g & b & 1 \end{array} \right]^T \cdot \left[ \begin{array}{cccc} r &  g & b & 1 \end{array} \right] \right),
\end{eqnarray}
where $triu(\cdot)$ is the vectorized form of the elements in the upper triangular of a matrix (to discard redundancies such as $r \cdot g$ and $g \cdot r$). The operator $W \in \mathbb{R}^{3 \times 10}$  maps the second-order monomials of each pixel to a new RGB value. 
This family of transformations has been selected by the observation that on real pairs of raw low-light and processed well-lit images, linear regression is inadequate for approximating the transformation between the two. On the other hand, a quadratic transformation produces pleasant looking images.

\changes{Our solution for the high-level stage of the network is related to few works \cite{gharbi2017deep,getreuer2017blade,jiang2017learning}, where a model is also learned to predict a transformation that is then applied to an input image. However, they use local transformation and not global like in our work.} \minorrev{The choice of global transformation is a limitation of our proposed architecture, preventing it from learning classical local tone mapping used for HDR. We found, however, that when combined with the low-level part of the network, which applies local additive corrections, the usage of a global model is sufficient for the task at hand and enjoys better convergence and stability.}

\reb{
In another recent work \cite{chen2018learning}, done in parallel to ours, it was also suggested learning to map a raw low-light image to a well-lit processed image. They used a U-Net architecture and avoided residual blocks.
}


\subsection{Loss}
A commonly used loss for image restoration is the $\ell_2$-distance. While it optimizes mean squared error (MSE), which is directly related to the peak signal-to-noise ratio (PSNR), it leads to inferior results with respect to perceptual quality compared to other loss functions \cite{zhao2017loss}. 

When training the network just for the task of joint denoising and demosaicing (i.e., using only its lower level), we use the $\ell_2$-loss and measure the performance in PSNR. Yet, in the case of the full ISP, PSNR is meaningless because a small deviation in the global color will result in a very large error (and low PSNR), while having no effect on perceptual quality. In this case, we use a combination of the $\ell_1$ norm and the multi scale structural similarity index (MS-SSIM) to get a higher perceptual quality as suggested in \cite{zhao2017loss}.

In the full ISP case, the loss function is defined in the Lab domain. Because the network operates in the RGB color space, for calculating the loss, the network output needs to go through an RGB-to-Lab color conversion operator. This operator is differentiable almost everywhere and it is easy to calculate its gradient.
While we compute the $\ell_1$ loss on all the three Lab channels, the MS-SSIM is evaluated only on the luminance (L) channel:
\begin{eqnarray}
Loss(\hat{I},I) &=& \left(1-\alpha\right) \|\mathrm{Lab}(\hat{I}) - \mathrm{Lab}(I)\|_1 \\ \nonumber && + 
{\alpha \, MSSSIM\left(\mathrm{L}(\hat{I}), \mathrm{L}(I)\right)}.
\end{eqnarray}
The reasoning behind this design choice is that we want the model to learn both local (captured by MS-SSIM) and global (enforced by $\ell_1$) corrections. Applying MS-SSIM to the luminance channel enables learning local luminance corrections even before the color (a and b channels) has converged to the target value.  Also, MS-SSIM is based on local statistics and is mostly affected by the higher frequency information, which is of lower significance in the color channels.

\section{Joint Denoising and Demosaicing}

\subsection{Evaluation}
Since for the full ISP it is hard to define objective metrics, we start by evaluating our solution on the task of joint denoising and demosaicing.  There is a considerable research examining this task and recent studies (e.g., \cite{gharbi2016deep} and \cite{klatzer2016learning}) have benchmarked on the MSR demosaicing dataset \cite{khashabi2014joint}.
\reb{
This dataset is generated by down-sampling a mosaiced image, so each pixel will have its ground truth red, green and blue values. The noise in this dataset is designed to be realistic, the level of noise is estimated in the original image and applied to the down-sampled image. We measured the standard deviation for the mosaiced noisy images compared to their corresponding ground truth values and found the STD range is $\sigma \in \left[1,10\right]$.
}
For the task of joint denoising and demosaicing, we used the Panasonic images in the MSR dataset for training, and report results for both the Panasonic and Canon test sets (disjoint from the training sets).

As the denoising and demosaicing task requires only local image modifications, we only use the low-level part of the network, \reb{ i.e. the output of the last residual block is used as the model output.
}
We set the number of blocks to $N_{ll}=20$. The mosaiced raw image is transformed to an RGB image by bilinear interpolation during the preprocessing stage. We retained the test set as specified in the dataset and split the remaining $300$ images into $270$ for training and $30$ for validation. The resolution of all images was $132 \times 220$; although some were captured in portrait mode and some in landscape mode, we used all images in landscape orientation. The data were further augmented with random horizontal and vertical flipping. The network was trained for $5000$ epochs using the Adam optimizer with learning rate $5 \times 10^{-5},\  \beta _1 = 0.9,\   \beta _2 = 0.999$ and $\epsilon = 10^{-8}$.

\begin{figure*}[tb]
    \centering
    \begin{tabular}{c@{\hskip 0.01\textwidth}c@{\hskip 0.01\textwidth}c}
        \includegraphics[width = 0.32\textwidth]{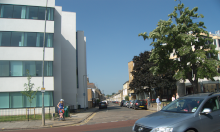} &
        \includegraphics[width = 0.32\textwidth]{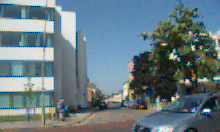} &
        \includegraphics[width = 0.32\textwidth]{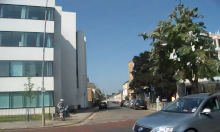} \\  
        
        \includegraphics[width = 0.32\textwidth]{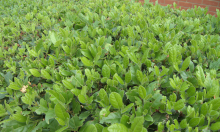} &
        \includegraphics[width = 0.32\textwidth]{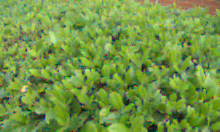} &
        \includegraphics[width = 0.32\textwidth]{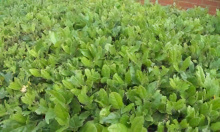} \\  
        
        \includegraphics[width = 0.32\textwidth]{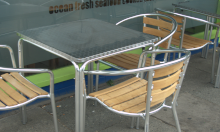} &
        \includegraphics[width = 0.32\textwidth]{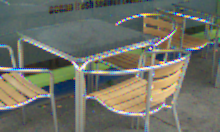} &
        \includegraphics[width = 0.32\textwidth]{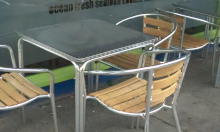} \\  
        
        {Ground Truth} &
        {Input (Bilinear)} &
        {Output} \\
    \end{tabular}      
    \smallskip 
    \caption{\small \textbf{Results for joint denoising and demosaicing.} Trained and tested on images from the MSR Dataset \cite{khashabi2014joint}. The input to the network is demosaiced by bilinear interpolation. The artifacts caused by the interpolation are visible in the middle column images and the model learns to remove them quite well.}
    \label{fig_msr_results}
\end{figure*}

\begin{figure*}[tb]
    \centering
    \begin{tabular}{cc}
        \includegraphics[width = 0.45\textwidth]{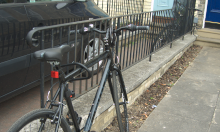} &
        \includegraphics[width = 0.45\textwidth]{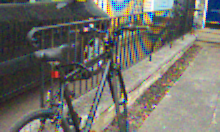} \\
        {Ground Truth} & {Input (Bilinear)} \\
        \includegraphics[width = 0.45\textwidth]{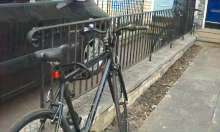} &
        \includegraphics[width = 0.45\textwidth]{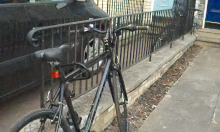} \\
         {SEM \cite{klatzer2016learning}} & {Ours} \\
    \end{tabular}      
    \smallskip 
    \caption[An example of handling Moir\'e artifacts]{\reb{An example of handling Moir\'e artifacts. Note the blue artifacts on the far building in SEM's \cite{klatzer2016learning} output that are perfectly removed by our model.}}
    \label{fig_moire}
\end{figure*}

Some visual examples of our restoration results are shown in Fig.~\ref{fig_msr_results}. \reb{A known challenge in demosaicing is the Moir\'e artifact, which is particularly observed in image locations with high frequency patterns. Figure \ref{fig_moire} demonstrates how well our method handles this artifact; it gets rid of it even in cases where competing methods fail (e.g., see the blue artifact on the far building in the SEM result, bottom left in Fig.~\ref{fig_moire}).}

Table~\ref{tab_msr_results} summarizes the  comparison to other methods on the MSR dataset. \reb{All numbers of the competing methods are taken from \cite{klatzer2016learning}}. Our proposed method achieves the best results for joint denoising and demosaicing on both the Panasonic and Canon test sets in the MSR dataset. Compared to the previous state-of-the-art results (SEM by \cite{klatzer2016learning}), our method produces an improvement of $0.38dB$ (linear space) and $0.72dB$ (sRGB space) on the Panasonic test set, and of $0.61dB$ (linear) $1.28dB$ (sRGB) on the Canon test set. 

This experiment corroborates the ability of our deep learning-based model to generalize well to a different dataset (training on Panasonic images and testing on Canon images). This strength of our solution is also noticeable with an improvement of $0.71dB/1.05dB$ we get over another deep learning base method \cite{gharbi2016deep} on Linear/sRGB Panasonic. 
\reb{Note that we train only on the MSR dataset, which contains a few hundred images, while the training procedure in \cite{gharbi2016deep} uses, in addition, an external dataset with millions of images for mining hard examples and training on them.}

\begin{table}[tb]
\centering
\small
  \begin{tabular}{lllll} 
    \toprule
    {}                                 &  \multicolumn{2}{c}{Panasonic} & \multicolumn{2}{c}{Canon}\\
    Method                             &   Linear     &   sRGB     &   Linear    &   sRGB    \\ 
   \midrule
    Matlab \cite{malvar2004high}     &   34.16   &   27.56    &    36.38    &    29.1    \\
    OSAP \cite{lu2010demosaicking}     &   36.25    &    29.93    &    39        &    31.95    \\
    WECD \cite{su2006highly}         &   36.51    &    30.29    &    -        &    -        \\
    NLM \cite{buades2009self}        &   36.55    &    30.56    &    38.82    &    32.28    \\
    DMMSE \cite{zhang2005color}        &   36.67    &    30.24    &    39.48    &    32.39    \\
    LPA    \cite{paliy2007spatially}    &   37        &    30.86    &    39.66    &    32.84    \\
    CS \cite{getreuer2009contour}    &   37.2    &    31.41    &    39.82    &    33.24    \\
    JMCDM \cite{chang2015color}        &   37.44    &    31.35    &    39.49    &    32.41    \\
    RTF    \cite{khashabi2014joint}     &   37.77    &    31.77    &    40.35    &    33.82    \\
    FlexISP \cite{heide2014flexisp} &   38.28    &    31.76    &    40.71    &    33.44    \\
    DJDD \cite{gharbi2016deep}        &   38.6    &    32.6    &    -        &    -        \\
    SEM     \cite{klatzer2016learning}    &   38.93    &    32.93    &    41.09    &    34.15    \\
    Ours                            & \textbf{39.31} & \textbf{33.65} &    \textbf{41.7} &    \textbf{35.43} \\
    \bottomrule
  \end{tabular}
  \vspace{2mm}
\caption{PSNR for Joint denoising and demosaicing of the MSR dataset. \reb{Other methods results are taken from \cite{klatzer2016learning}.}} \label{tab_msr_results}
\end{table}

\subsection{\reb{Choosing hyper-parameters}}
\begin{figure}[tbh]
    \centering
        \includegraphics[width = 0.85\linewidth]{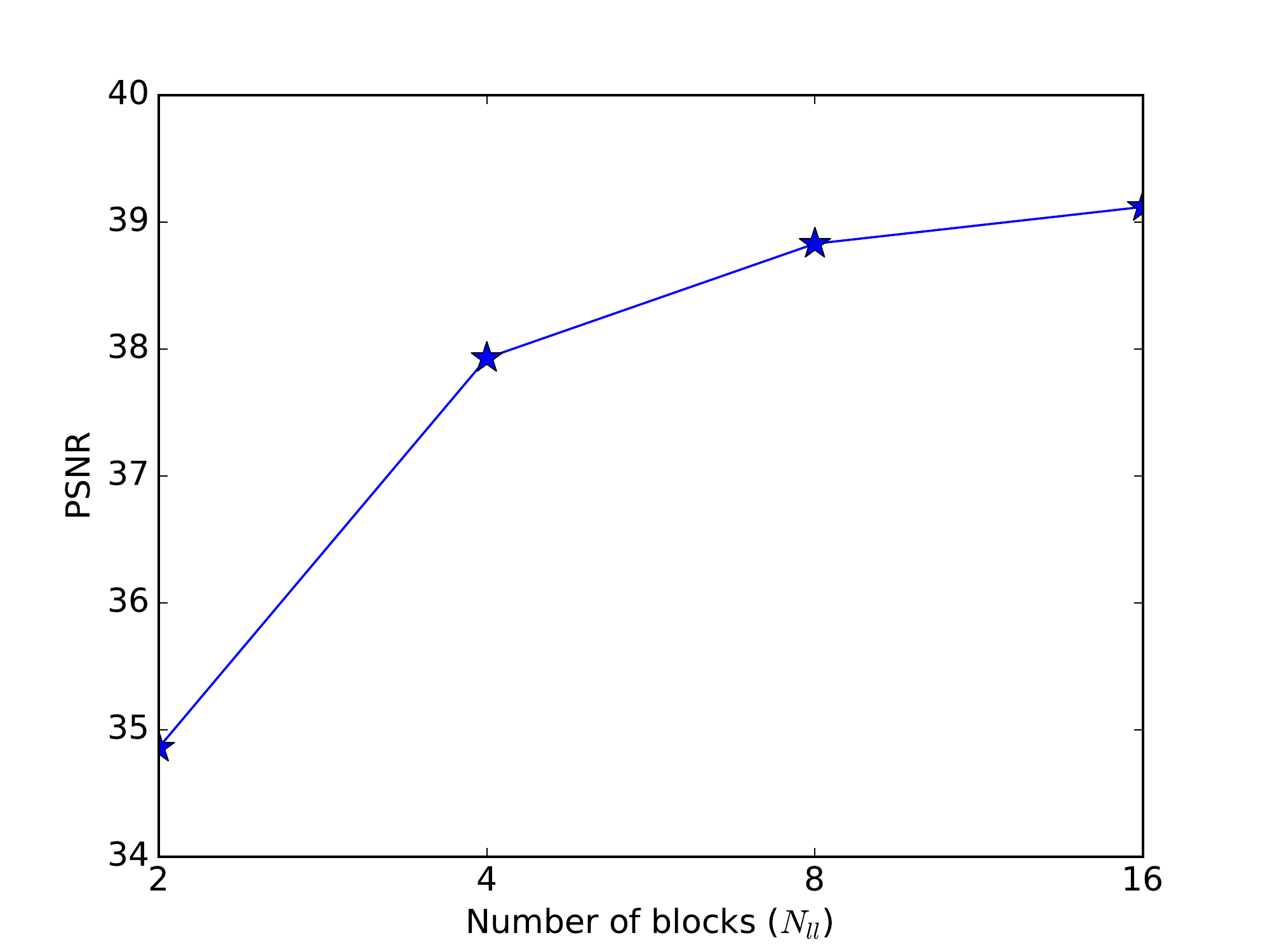}   
    \smallskip 
    \caption{\reb{PSNR vs. number of residual blocks}}
    \label{fig_psnr_layers}
\end{figure}

\reb{
The large number of residual blocks used in our network has two main effects. First, it allows the network to learn more complicated functions with more parameters and more nonlinear units. Second, it generates larger receptive field, i.e. for $N$ $3\times3$ convolution layers each pixel at the output is a function of $2N+1 \times 2N+1$ neighboring input pixels. Figure \ref{fig_psnr_layers} shows PSNR performance as a function of the number of residual blocks. As expected, we observe increased performance for deeper network, reaching convergence, or diminishing returns, at around 16 layers.} 

\begin{figure}[tbh]
    \centering
        \includegraphics[width = 0.85\linewidth]{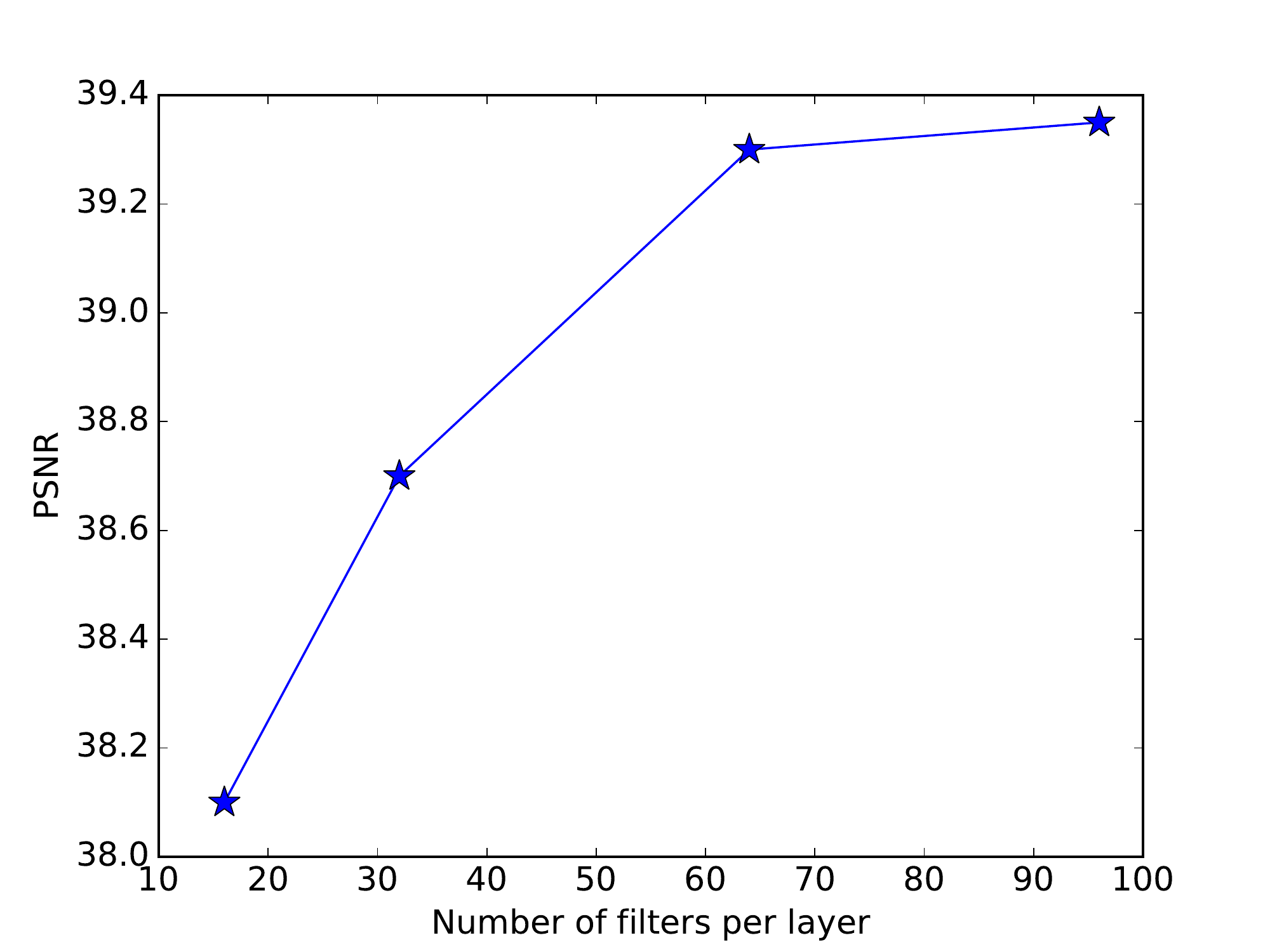}   
    \smallskip 
    \caption{\reb{PSNR vs. number of filters per layer}}
    \label{fig_psnr_filters}
\end{figure}

\reb{
The number of filters per layer affects the expressiveness of the model. Figure \ref{fig_psnr_filters} shows how (for a network with 20 layers) performance increases with more filters . Convergence is reached  at around 64 filters per layer. Note that increasing the number of filters by a factor $a$ results in a factor $a^2$ in the number of parameters, while the parameters number scales only linearly with number of layers.
}

\subsection{\reb{Effect of skip connections}}
\reb{
Training very deep networks has problems with convergence due to vanishing or exploding gradients. Vanishing gradients are caused when the gradient of the loss with respect to a parameter is too small to have any effect. Exploding gradients is the result of accumulated error in the calculation of the update step. Both are more apparent in very deep networks because there is a long path of layers between the loss and the first layers of the network, which implies many multiplications in the backward pass that are very likely to either converge to zero or explode.
}

\begin{figure}[tbh]
    \centering
        \includegraphics[width = 1\linewidth]{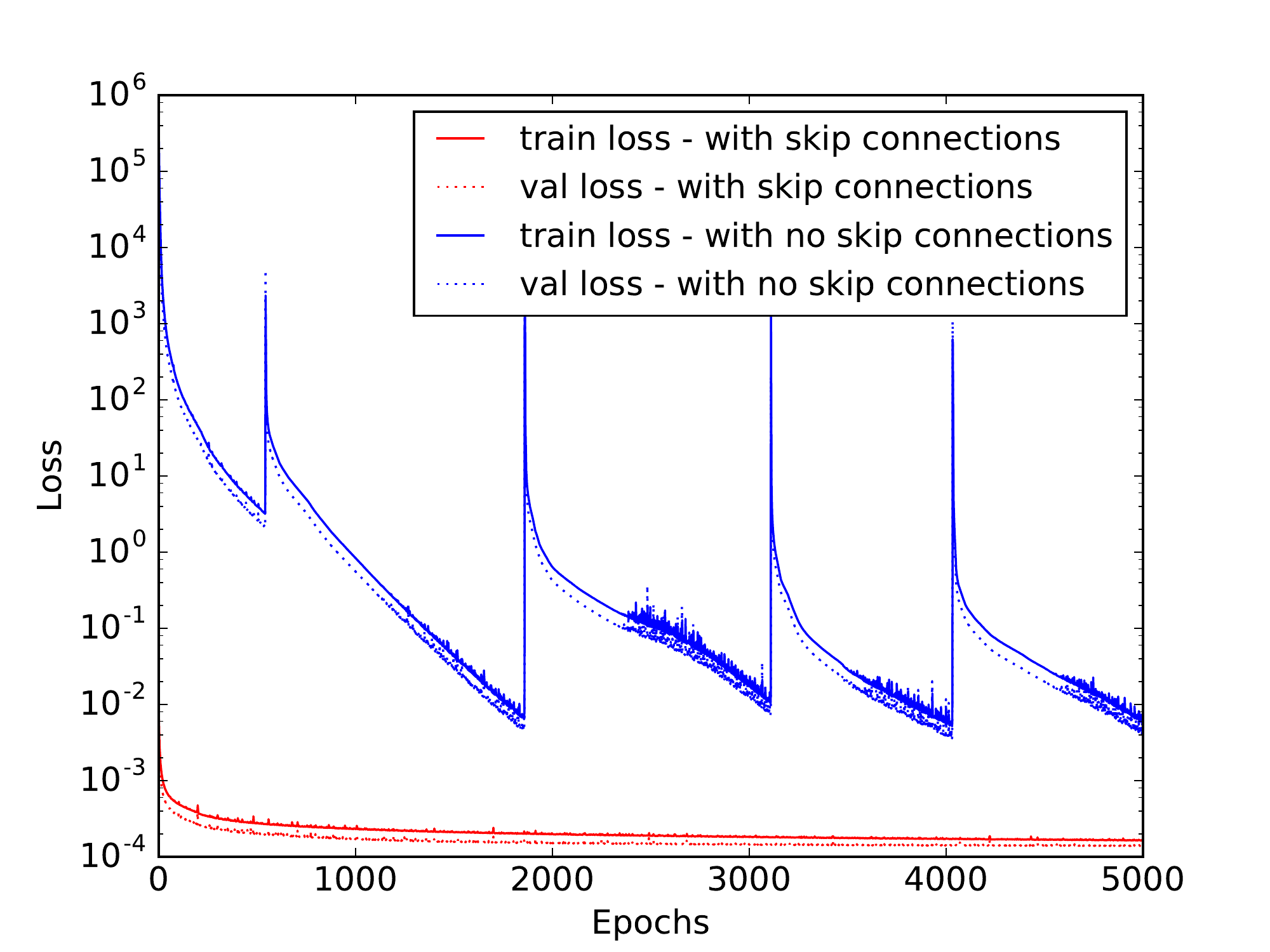}   
    \smallskip 
    \caption[Training with and without skip connection]{\reb{Training with and without skip connection. Without skip connections the training is not stable and after 5000 epochs reaches a loss 2 order of magnitude higher than the same model with skip connections}}
    \label{fig_skip_connection}
\end{figure}

\reb{
Skip connections, or "residual blocks", were suggested in \cite{He2015} as a way of having shorter paths from the output of the network to the first layers. These blocks that compute the  residual features have been proven to be very successful for classification models. The same intuition holds for using residual blocks for regression networks, as used in our model. To show the importance of skip connections we trained a model where the skip connections have been removed. Fig.~\ref{fig_skip_connection} shows that the training of this model is unstable and its inability to converge to anything similar to the original model with skip connections.
}

\section{Full ISP}
\subsection{S7-ISP Dataset}
To assess the performance of our full pipeline, we generated a dataset of real-world images. For this purpose, we captured, with a Samsung S7 rear camera, different scenes using a special Android application that was developed to capture a sequence of images while the phone is on a tripod and without having to touch it (to avoid camera movement).

While the scenes were chosen to contain minimal motion, a total lack of motion during the acquisition process cannot be guaranteed because the capturing was not performed in a lab setting. For each scene, we captured a JPEG image using the camera fully automatic mode and saved the original raw image too. In addition, we captured a low-light image of the same scene, stored in both JPEG and raw formats. The low-light image was emulated by capturing the same scene with the exact same settings as those chosen by the camera in the automatic mode, except the exposure time that was set to be quarter of the automatic setting. Since the camera supports only a discrete set of predefined exposure times, the closest supported value was selected. 

\begin{figure*}[tb]
    \centering
        \includegraphics[width =1\textwidth]{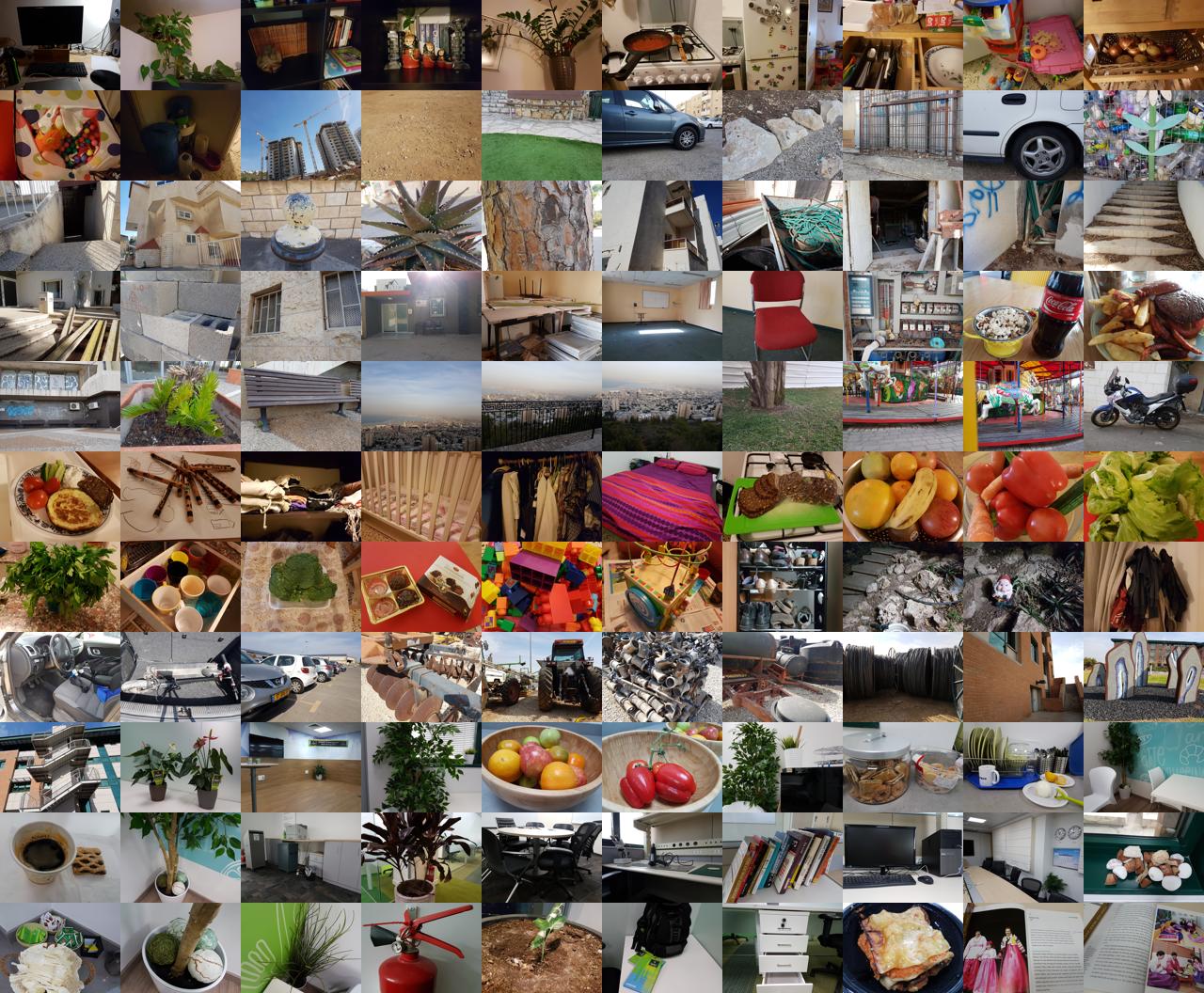}
    \caption{\reb{Captured scenes from the S7-ISP dataset}}
    \label{fig_dataset_thumbnails}
\end{figure*}

A total of $110$ scenes were captured and split to $90$, $10$ and $10$ for the training, validation and test sets, respectively. The relatively small number of images in the dataset is compensated by their $3024 \times 4032$ ($12M$ pixel) resolution. Thus, when training on patches, as common in DL-based methods, this dataset effectively contains many different samples. Even for relatively large $256 \times 256$ patches, it effectively contains over $20$ thousand of non-overlapping patches (and more than a billion different patches). 
\reb{
The scenes captured include indoors and outdoors, sun light and artificial light. Thumbnails of the scenes are displayed in Fig. \ref{fig_dataset_thumbnails}.}\footnote{The dataset is available at the \href{https://elischwartz.github.io/DeepISP}{project page}.}

\subsection{Mean Opinion Score}
\begin{figure}[tbh] 
\begin{center}
   \includegraphics[width=0.8\linewidth]{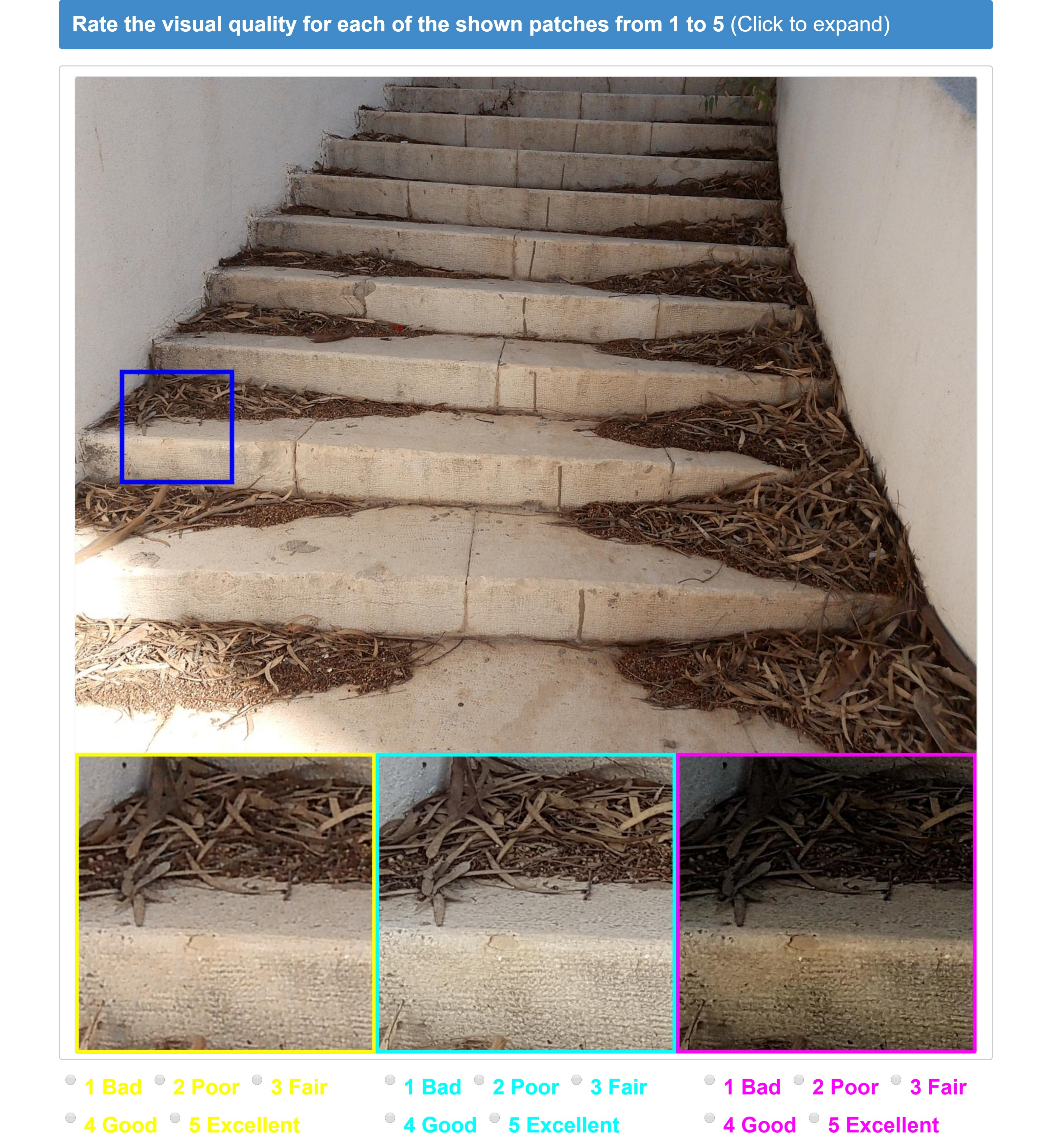}
\end{center}
       \caption{User interface presented to Amazon Mechanical Turk's workers for rating image patches}
    \label{fig_mech_turk_ui}
\end{figure}

To account for the fact that it is difficult to define an objective metric for the full pipeline, we performed a subjective evaluation, generating the mean opinion score (MOS) for each image using Amazon Mechanical Turk to quantitatively assess its quality. Two types of experiments were performed. The first experiment involved full images, where human evaluators were presented with a single image and were asked to rate its quality on a scale from 1 (bad) to 5 (excellent). In the second experiment, for rating the quality of details, evaluators were presented with multiple versions of the same patch side by side and were asked to rate each of them. The displayed patch size was set to $512 \times 512$ pixels (about $2\%$ of the total image). Evaluators were instructed to "rate the image quality according to factors like natural colors, details and sharpness (not the content of the image, e.g., composition)". Each evaluator was provided the opportunity to rate a specific example only once, but the exact same evaluators did not rate all examples. The user interface presented to the evaluators is shown in Fig. \ref{fig_mech_turk_ui}.

In addition to scoring by humans, we also evaluated image quality by a learned model from \cite{bosse2016deep} that was trained to estimate human evaluations. The model output was normalized to the range $[1,5]$. 

\subsection{DeepISP Evaluation}
The proposed end-to-end model was tested on the challenging task of learning the mapping between low-light raw input images to well-lit JPEG images (produced by the Samsung S7 ISP in automatic setting). 
The mosaiced raw image was transformed to RGB by bilinear interpolation as a preprocessing stage.
We use the proposed architecture with $N_{ll}=15$ and $N_{hl}=3$. 
For the MS-SSIM part of the loss, we used patches of $5\times5$ at two scales. 
The network was trained with a batch containing a single $1024 \times 1024$ patch cropped at random at each epoch from one of the training images. 
The data were augmented with random horizontal flipping. 
The training lasted for $700$ epochs using the ADAM optimizer with the following parameters: a learning rate of $5 \times 10^{-5},\  \beta _1 = 0.9,\   \beta _2 = 0.999,\   \epsilon = 10^{-8}$.

For faster convergence, the parameters of the learned operator $W$  were initialized with an affine operator $W_{init} \in \mathbb{R}^{3 \times 4}$. In this initialization, $W_{init}$, we mapped only the first-order monomials of each pixel to a new RGB value, so it did not contain the elements that correspond to second-order monomials (they were initialized to zero in $W$). We performed linear regression from input pixels to target pixels for each sample in the training set to get such an affine operator. As we get several operators in this way, $W_{init}$ was set to the average of them. We use a linear transformation $W_i \in \mathbb{R}^{3 \times 4}$ as the initialization of the full operator $W \in \mathbb{R}^{3 \times 10}$, zeroing its non-linear coefficients, due to this averaging operation. Unlike the affine operator, an average of multiple full operators did not lead to a reasonable operator, i.e., this average of several transforms in $\mathbb{R}^{3 \times 10}$ did not generate plausible images and did not serve as a good starting point for the optimization.

\minorrev{
It is important to note that a real camera ISP should be able to deal with motion artifacts, which are missing in this dataset. However, learning to generate high quality images with a shorter exposure time can help mitigating such artifacts.
}

\begin{figure}[tb]
\begin{center}
   \includegraphics[width=1.0\linewidth]{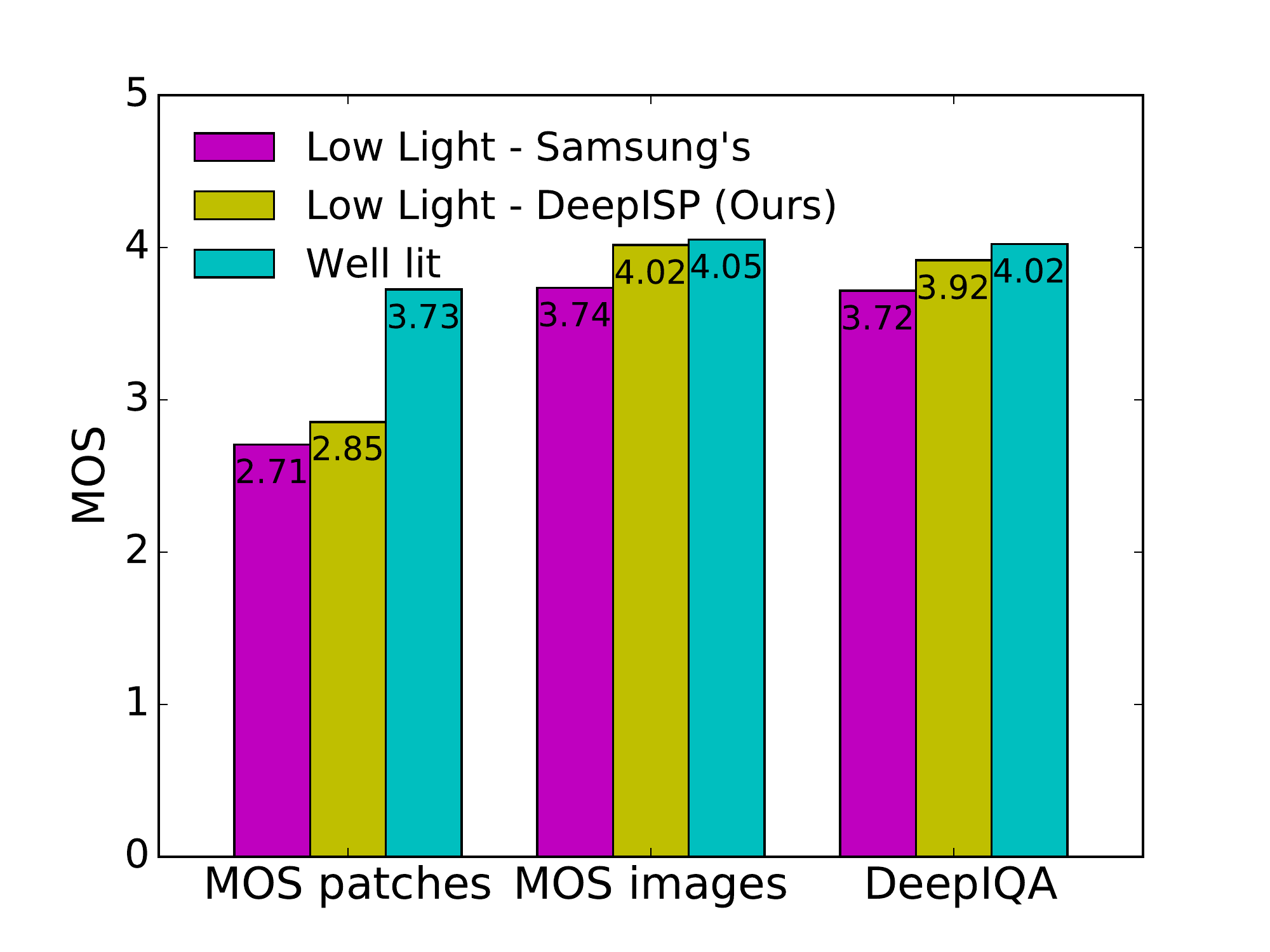}
\end{center}
       \caption{\textbf{MOS results for DeepISP.} Left: average human rating scores for random $512 \times 512$ patches. 
    Middle: average human rating scores of the full images. 
    Right: rating generated by a deep learning-based method for evaluating image quality \cite{bosse2016deep}. 
    Magenta: Samsung S7 ISP output for low-light images. 
    Yellow: our output for the same raw image. 
    Cyan: Samsung S7 ISP output for a well-lit scene, which serves as a ground truth.}
       \label{fig_isp_dataset_results_bar_graph}
    \label{fig:onecol}
\end{figure}

\begin{figure*}[tb]   
    \centering  
    \begin{tabular}{@{\hskip 0.005\textwidth}l@{\hskip 0.01\textwidth}l@{\hskip 0.01\textwidth}l@{\hskip 0.01\textwidth}l}
        \includegraphics[width = 0.24\textwidth]{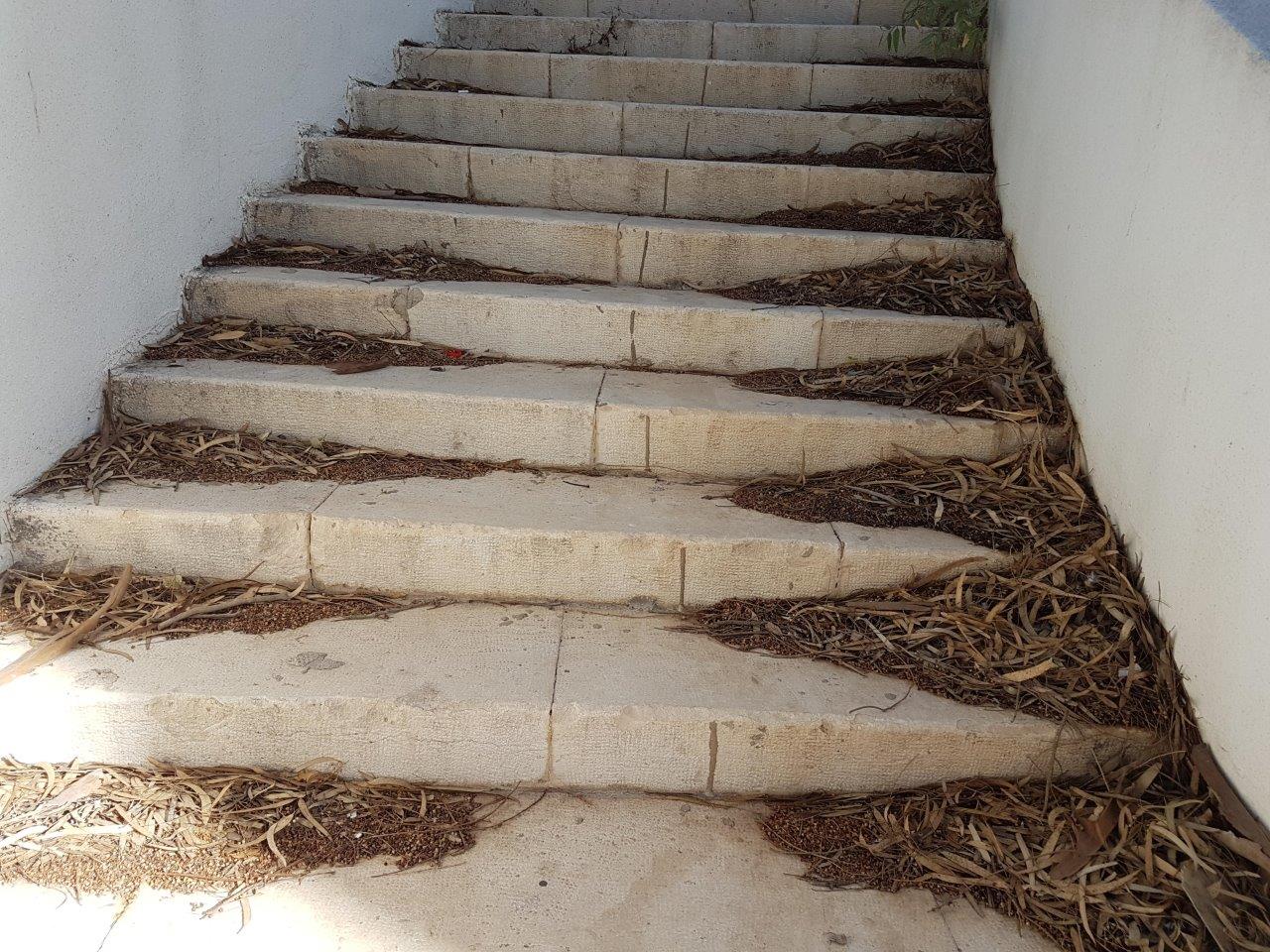} &
        \includegraphics[width = 0.24\textwidth]{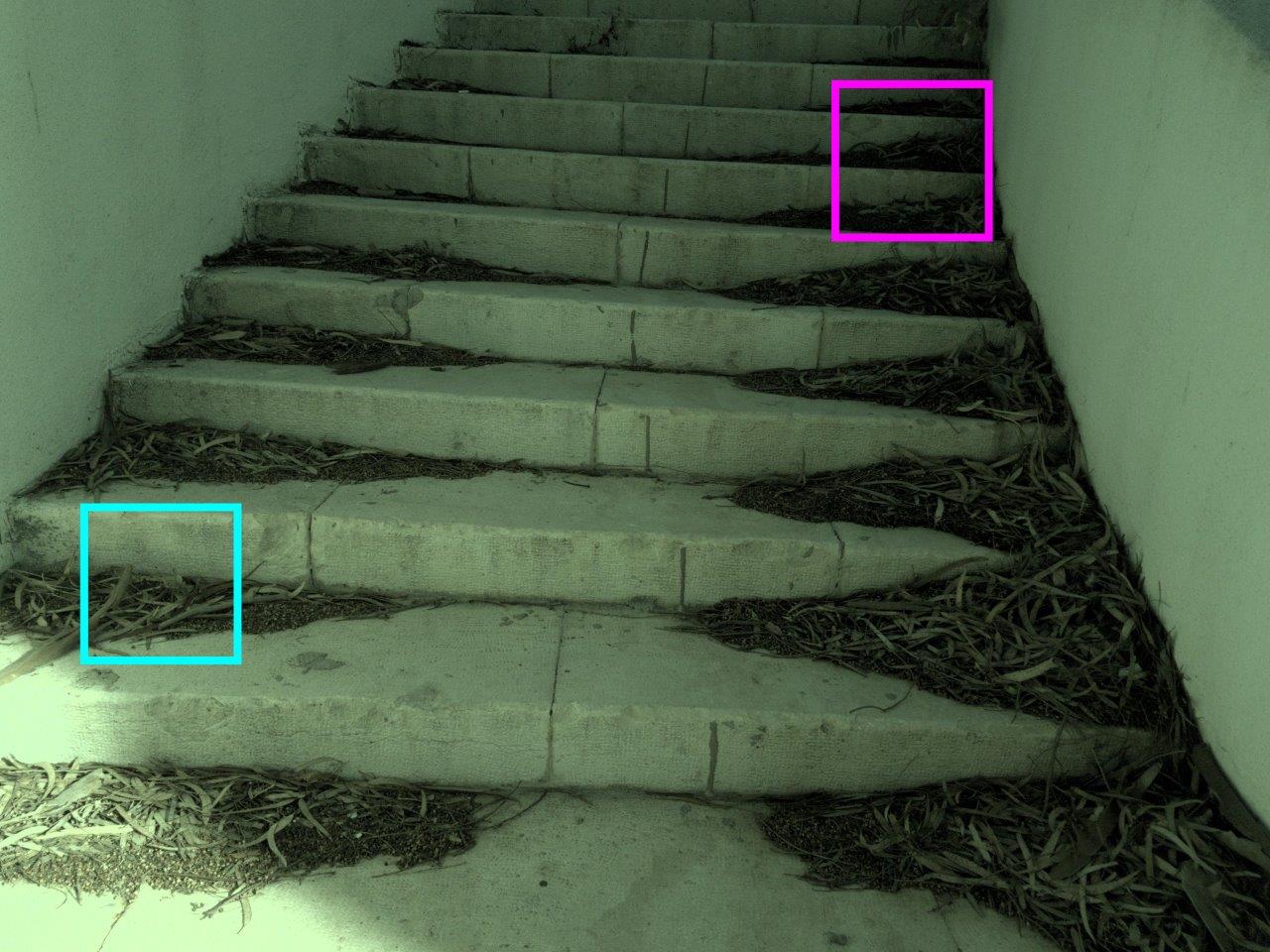} &
        \includegraphics[width = 0.24\textwidth]{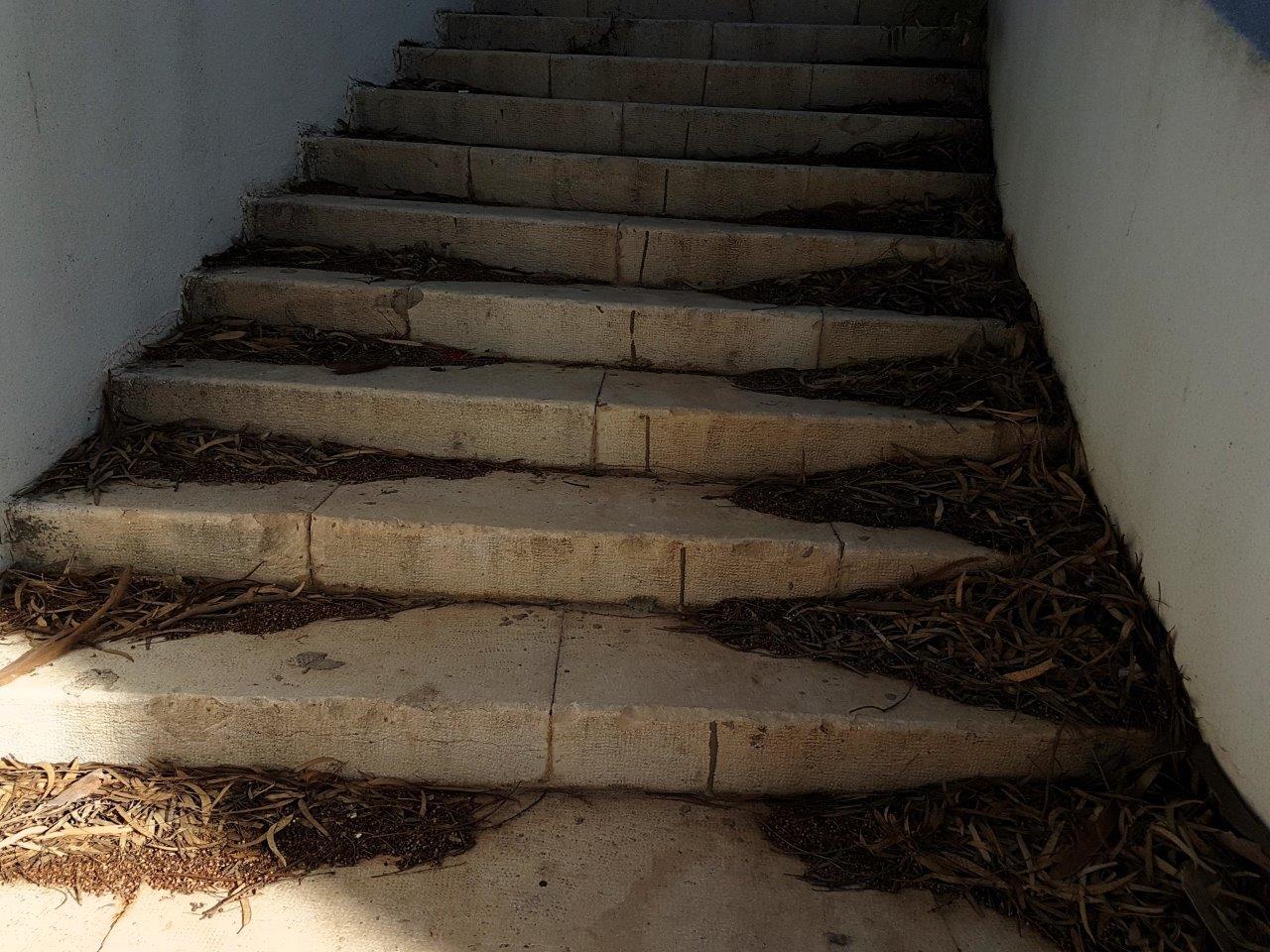} &
        \includegraphics[width = 0.24\textwidth]{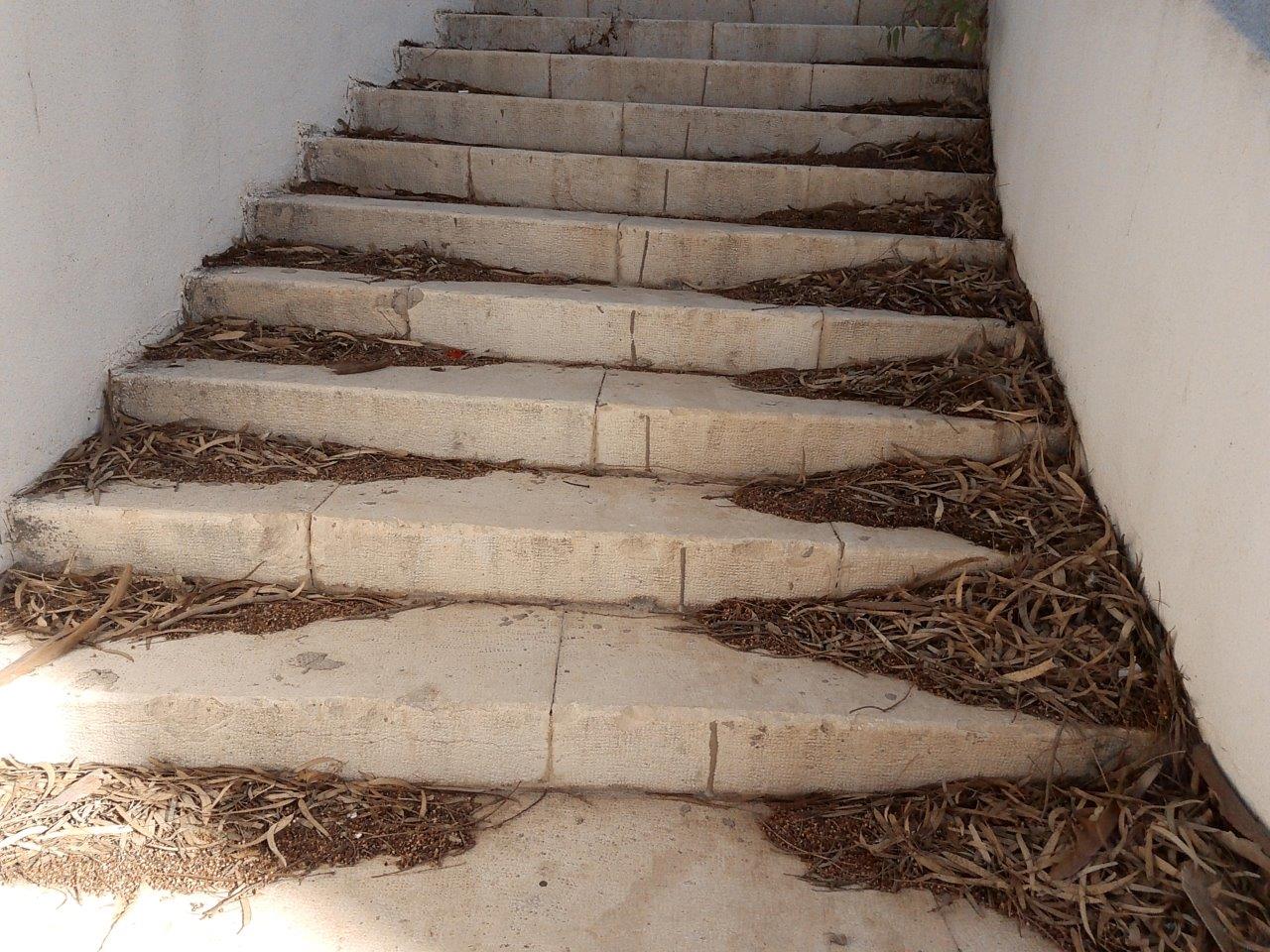} \\       
    \end{tabular}         
    \begin{tabular}{@{\hskip 0.005\textwidth}l@{\hskip 0.01\textwidth}l@{\hskip 0.01\textwidth}l@{\hskip 0.01\textwidth}l@{\hskip 0.01\textwidth}l@{\hskip 0.01\textwidth}l@{\hskip 0.01\textwidth}l@{\hskip 0.01\textwidth}c}   
          \includegraphics[width = 0.115\textwidth]{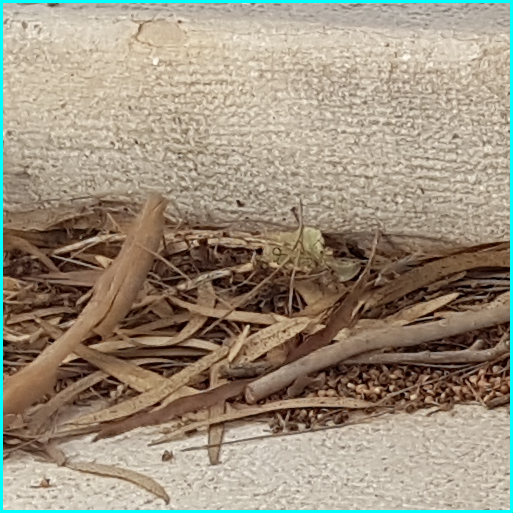} &
          \includegraphics[width = 0.115\textwidth]{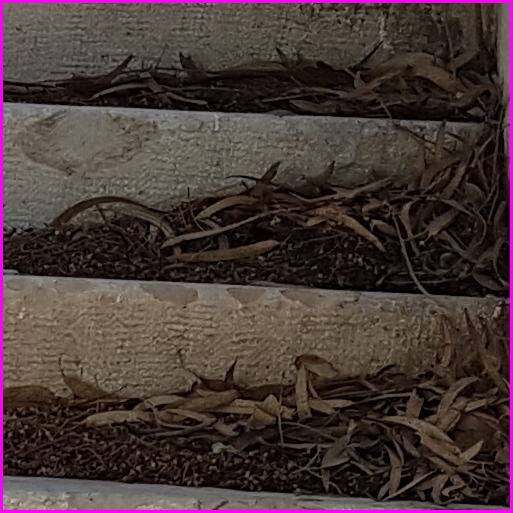} &
          \includegraphics[width = 0.115\textwidth]{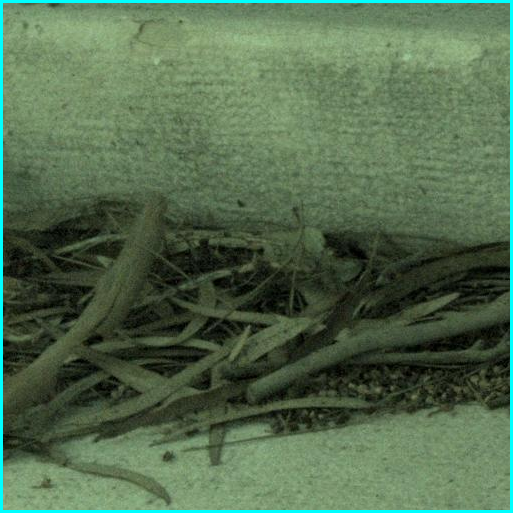} &
          \includegraphics[width = 0.115\textwidth]{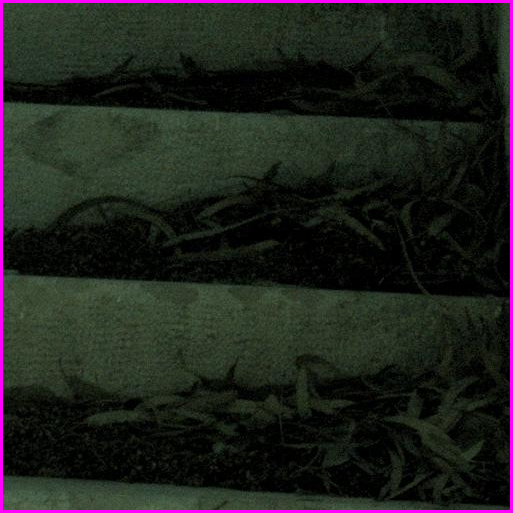} &
          \includegraphics[width = 0.115\textwidth]{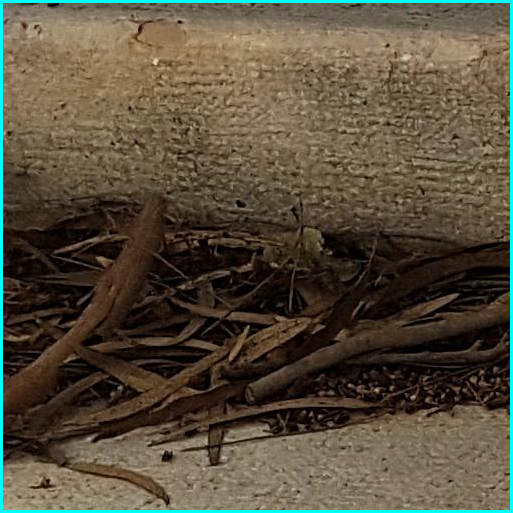} &
          \includegraphics[width = 0.115\textwidth]{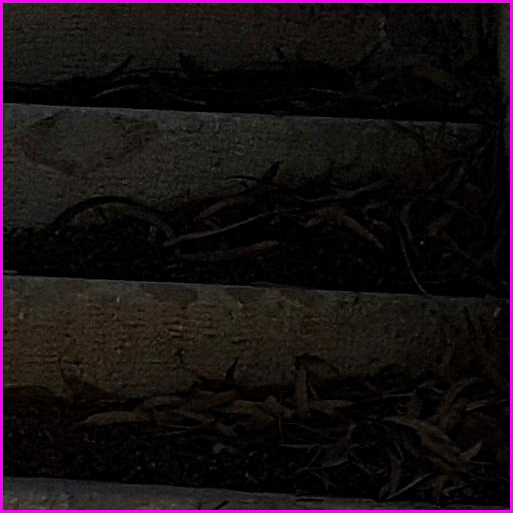} &
          \includegraphics[width = 0.115\textwidth]{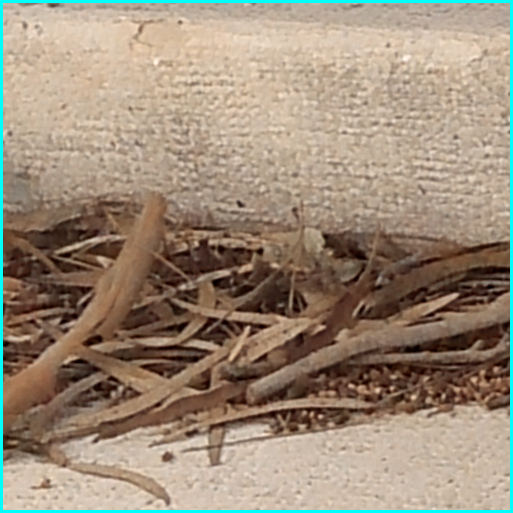} &
          \includegraphics[width = 0.115\textwidth]{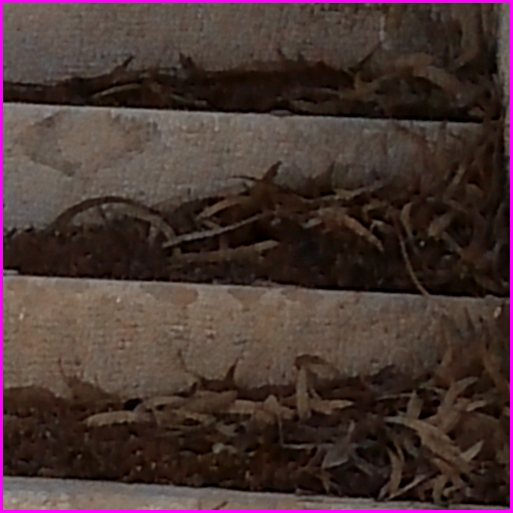} \\    
    \end{tabular}

    \begin{tabular}{@{\hskip 0.005\textwidth}l@{\hskip 0.01\textwidth}l@{\hskip 0.01\textwidth}l@{\hskip 0.01\textwidth}l}
        \includegraphics[width = 0.24\textwidth]{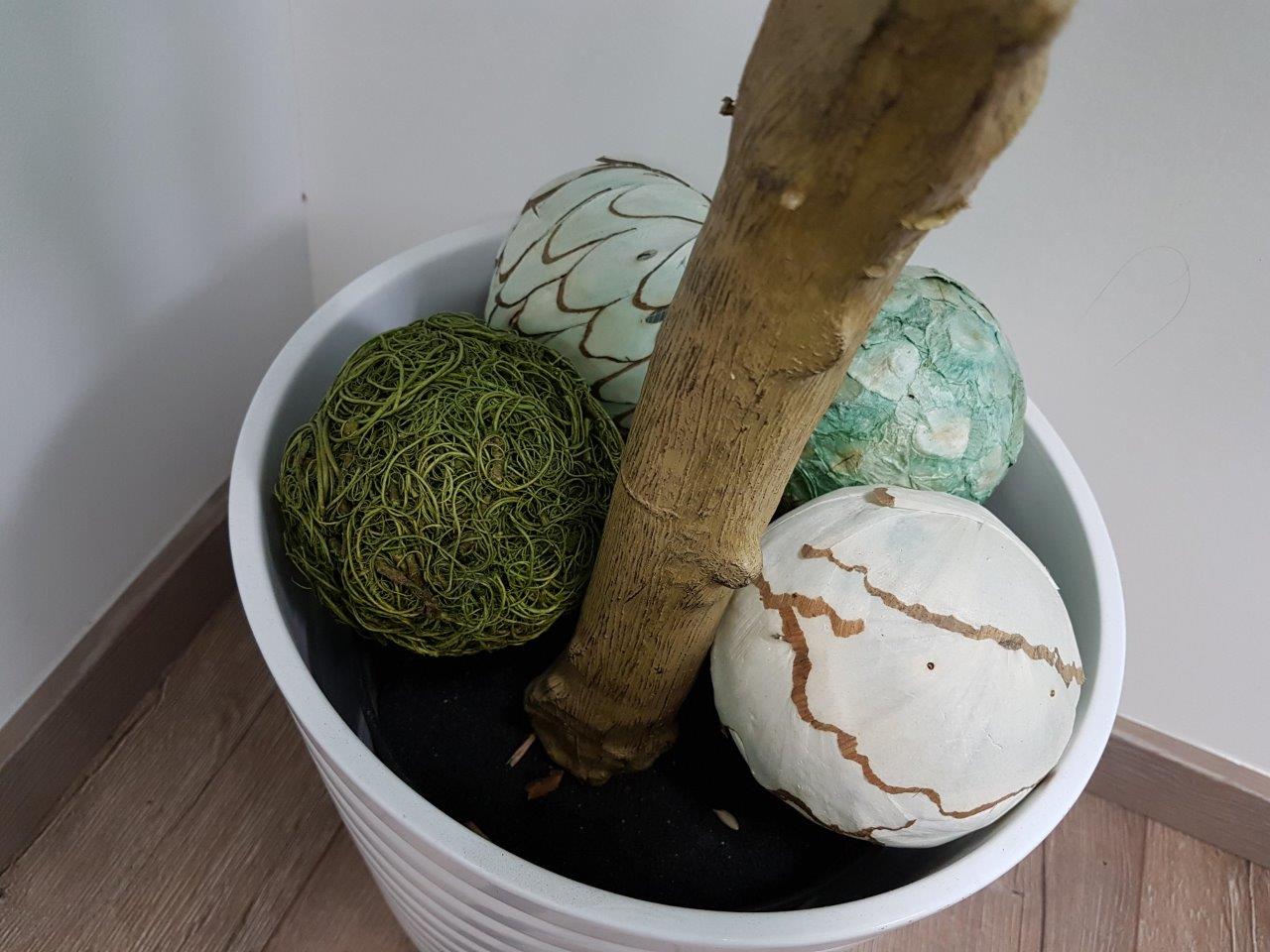} &
        \includegraphics[width = 0.24\textwidth]{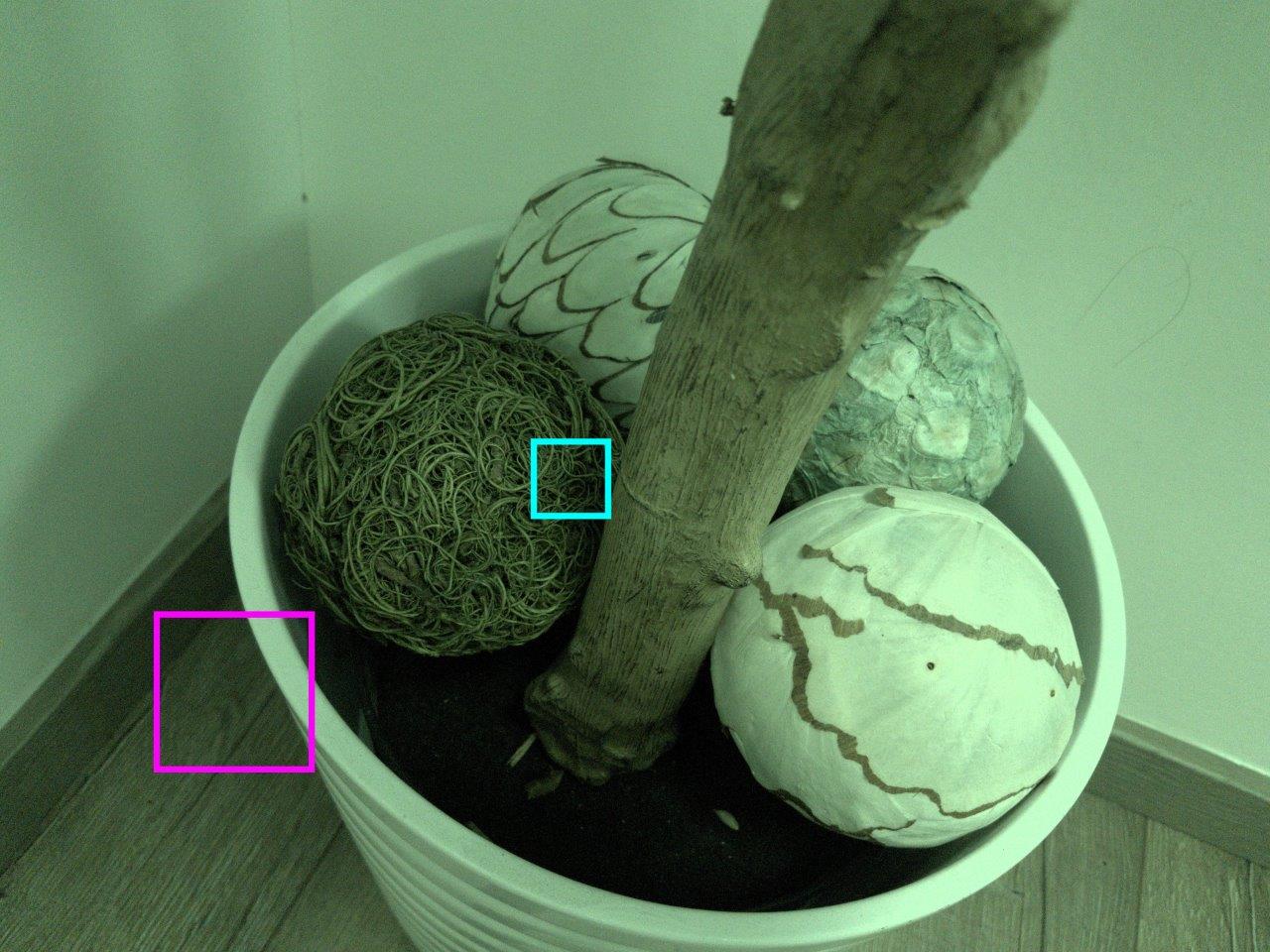} &
        \includegraphics[width = 0.24\textwidth]{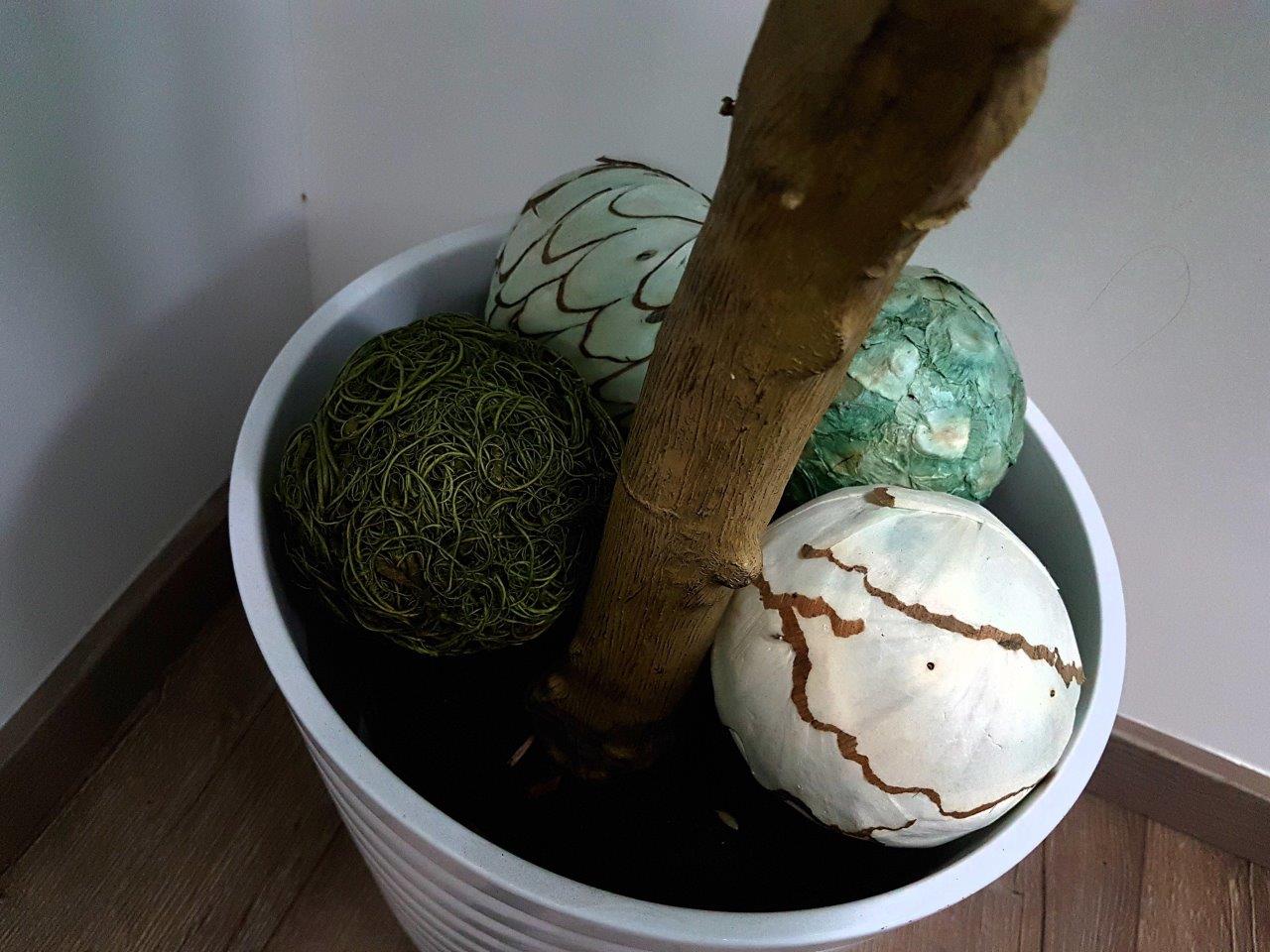} &
        \includegraphics[width = 0.24\textwidth]{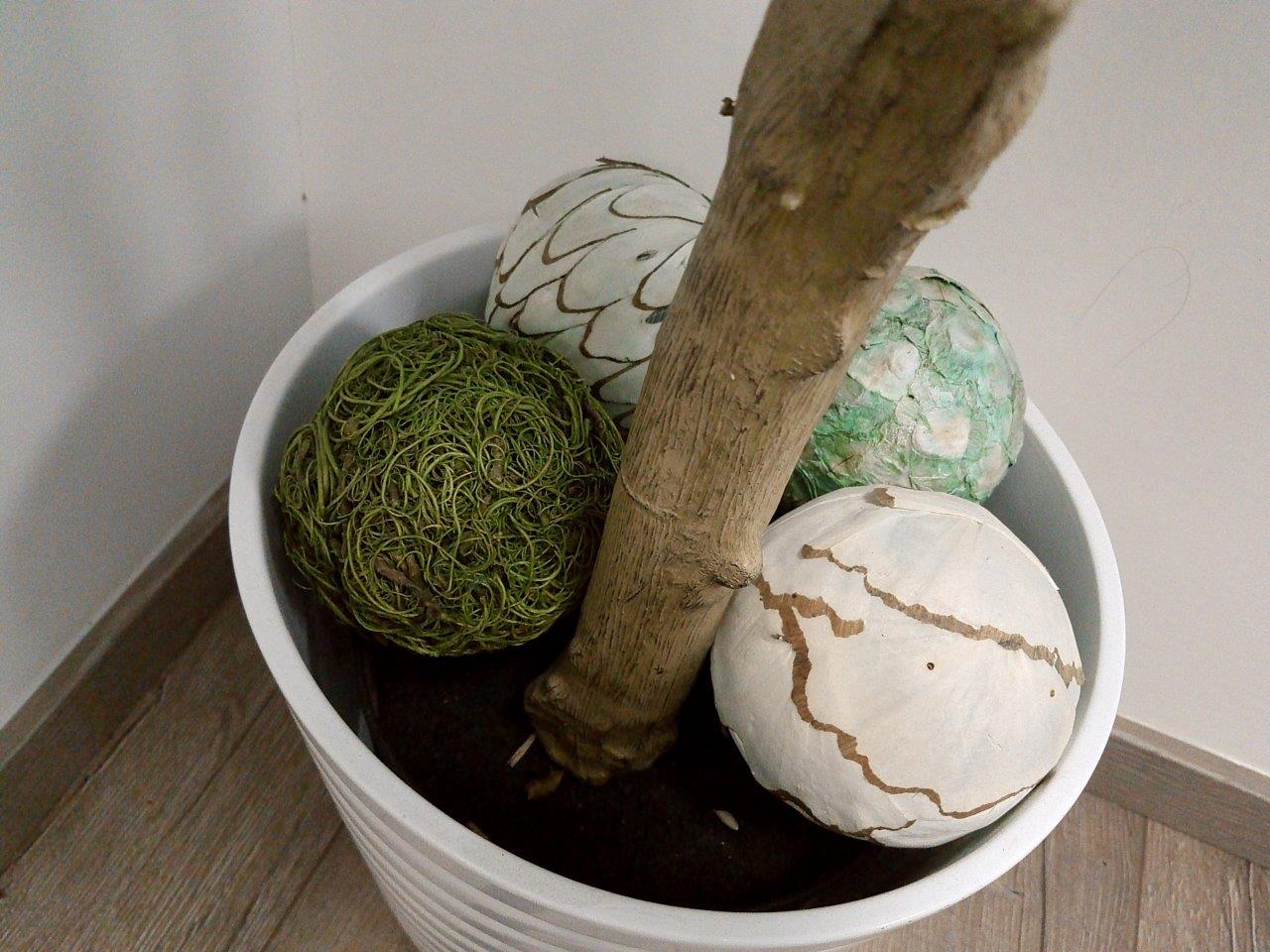} \\       
    \end{tabular}         
    \begin{tabular}{@{\hskip 0.005\textwidth}l@{\hskip 0.01\textwidth}l@{\hskip 0.01\textwidth}l@{\hskip 0.01\textwidth}l@{\hskip 0.01\textwidth}l@{\hskip 0.01\textwidth}l@{\hskip 0.01\textwidth}l@{\hskip 0.01\textwidth}c}   
          \includegraphics[width = 0.115\textwidth]{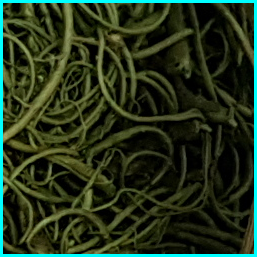} &
          \includegraphics[width = 0.115\textwidth]{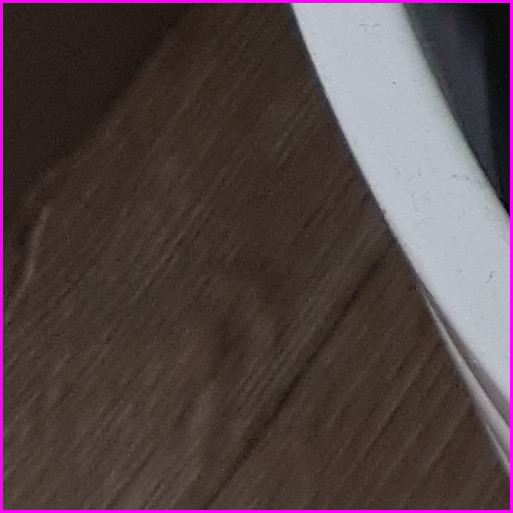} &
          \includegraphics[width = 0.115\textwidth]{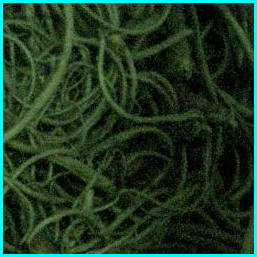} &
          \includegraphics[width = 0.115\textwidth]{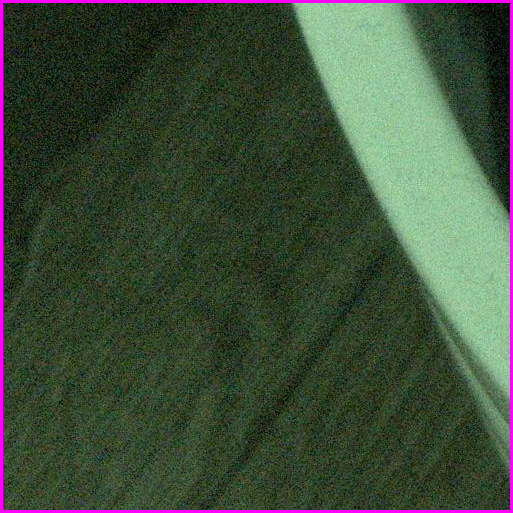} &
          \includegraphics[width = 0.115\textwidth]{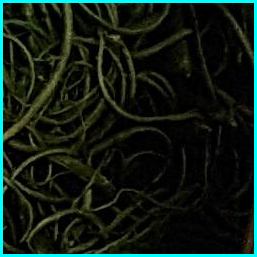} &
          \includegraphics[width = 0.115\textwidth]{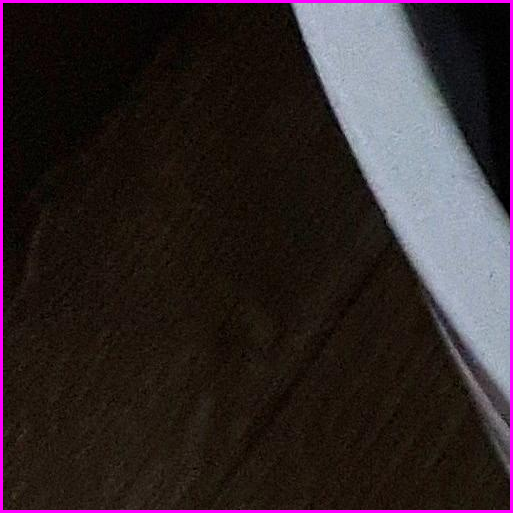} &
          \includegraphics[width = 0.115\textwidth]{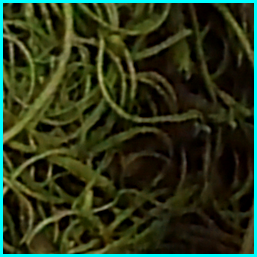} &
          \includegraphics[width = 0.115\textwidth]{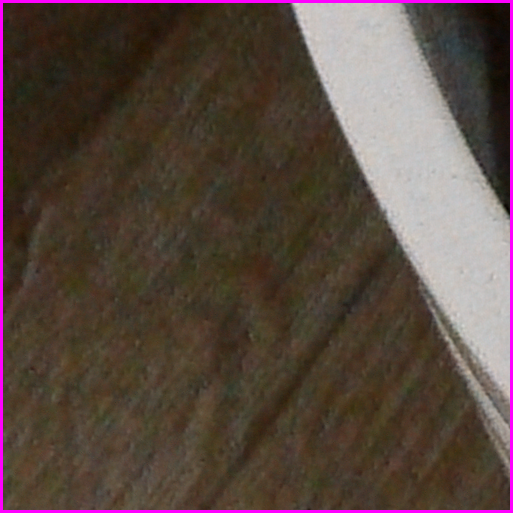} \\    
    \end{tabular}
    
    \begin{tabular}{@{\hskip 0.005\textwidth}l@{\hskip 0.01\textwidth}l@{\hskip 0.01\textwidth}l@{\hskip 0.01\textwidth}l}
        \includegraphics[width = 0.24\textwidth]{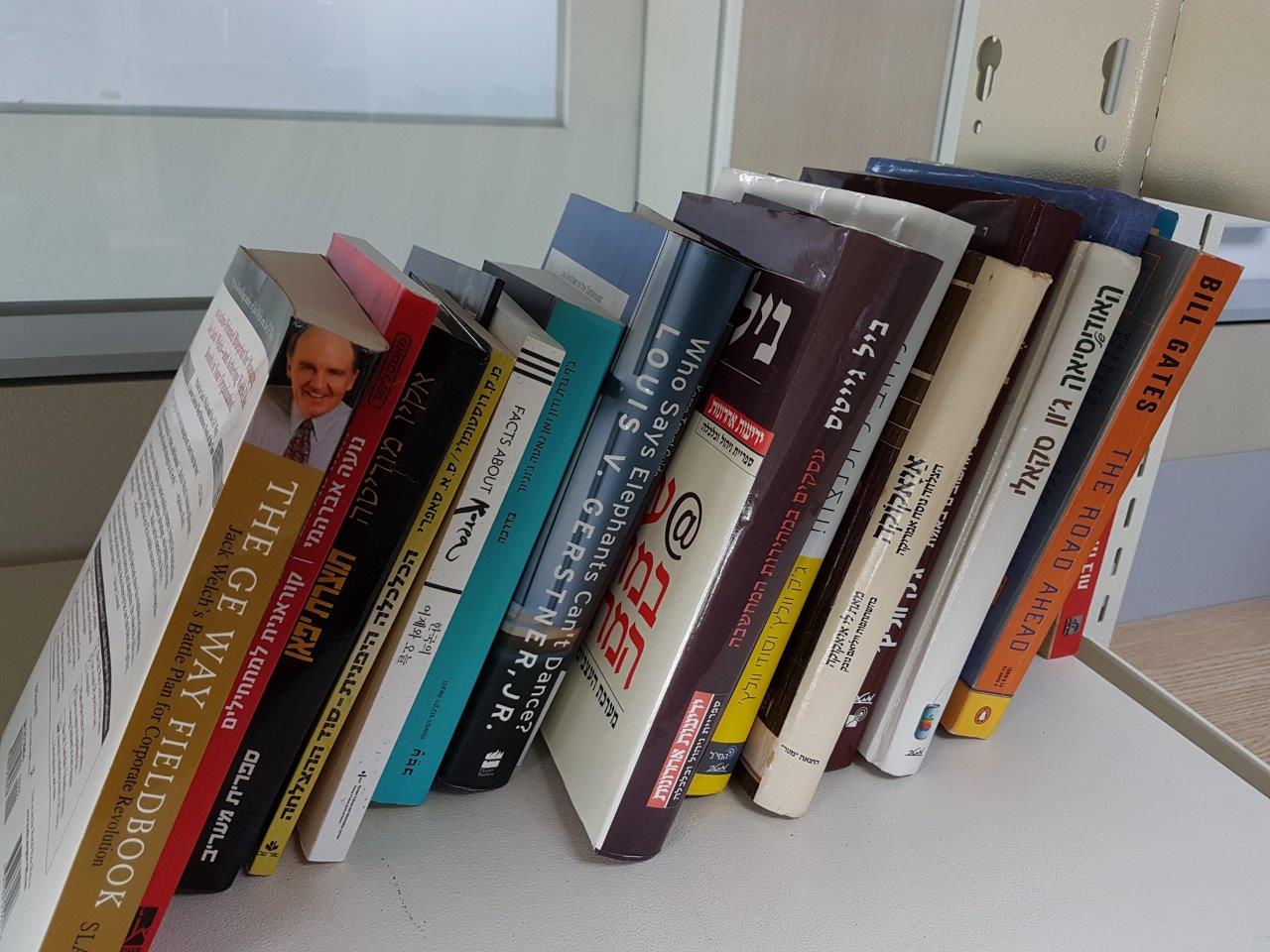} &
        \includegraphics[width = 0.24\textwidth]{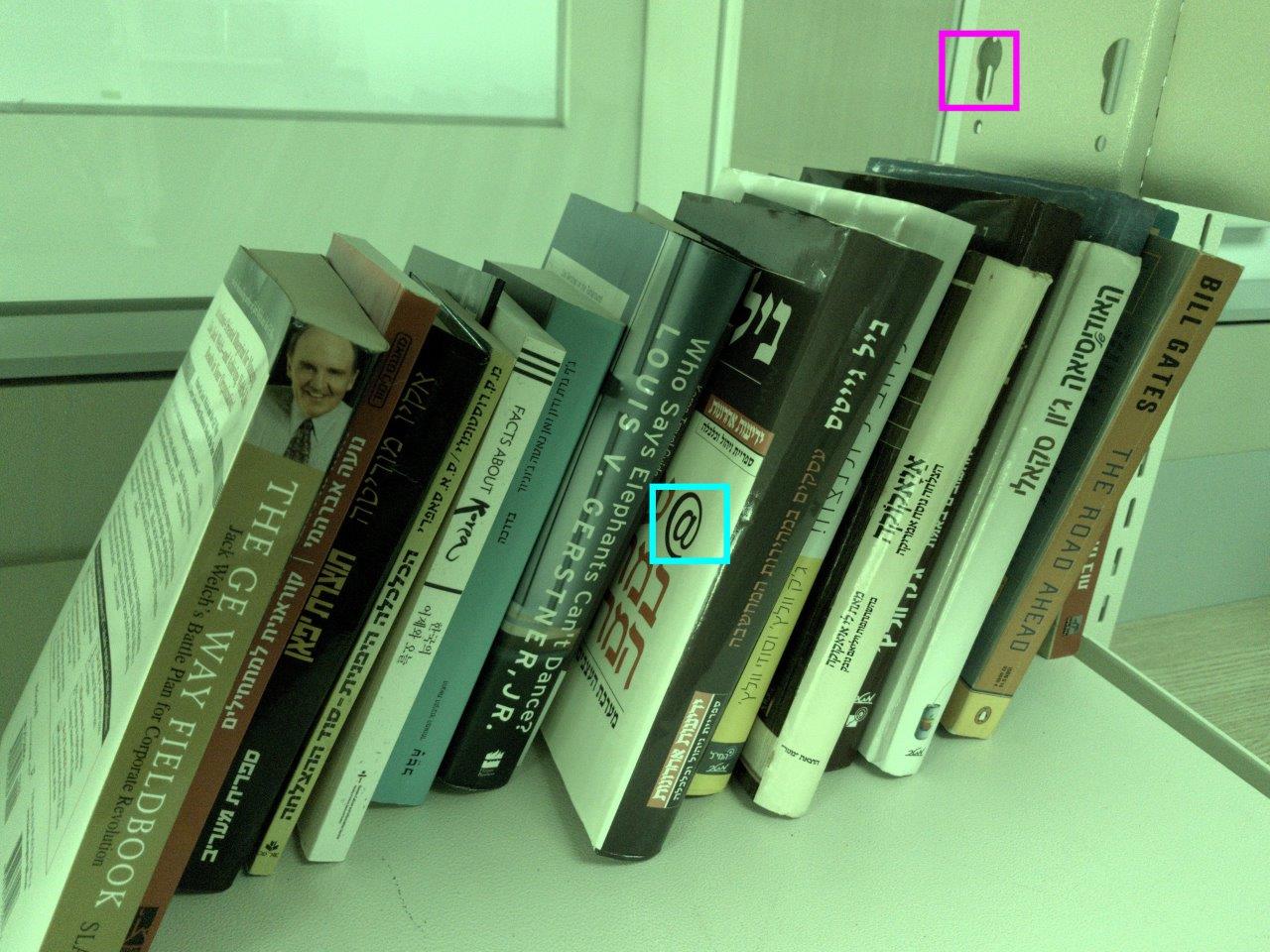} &
        \includegraphics[width = 0.24\textwidth]{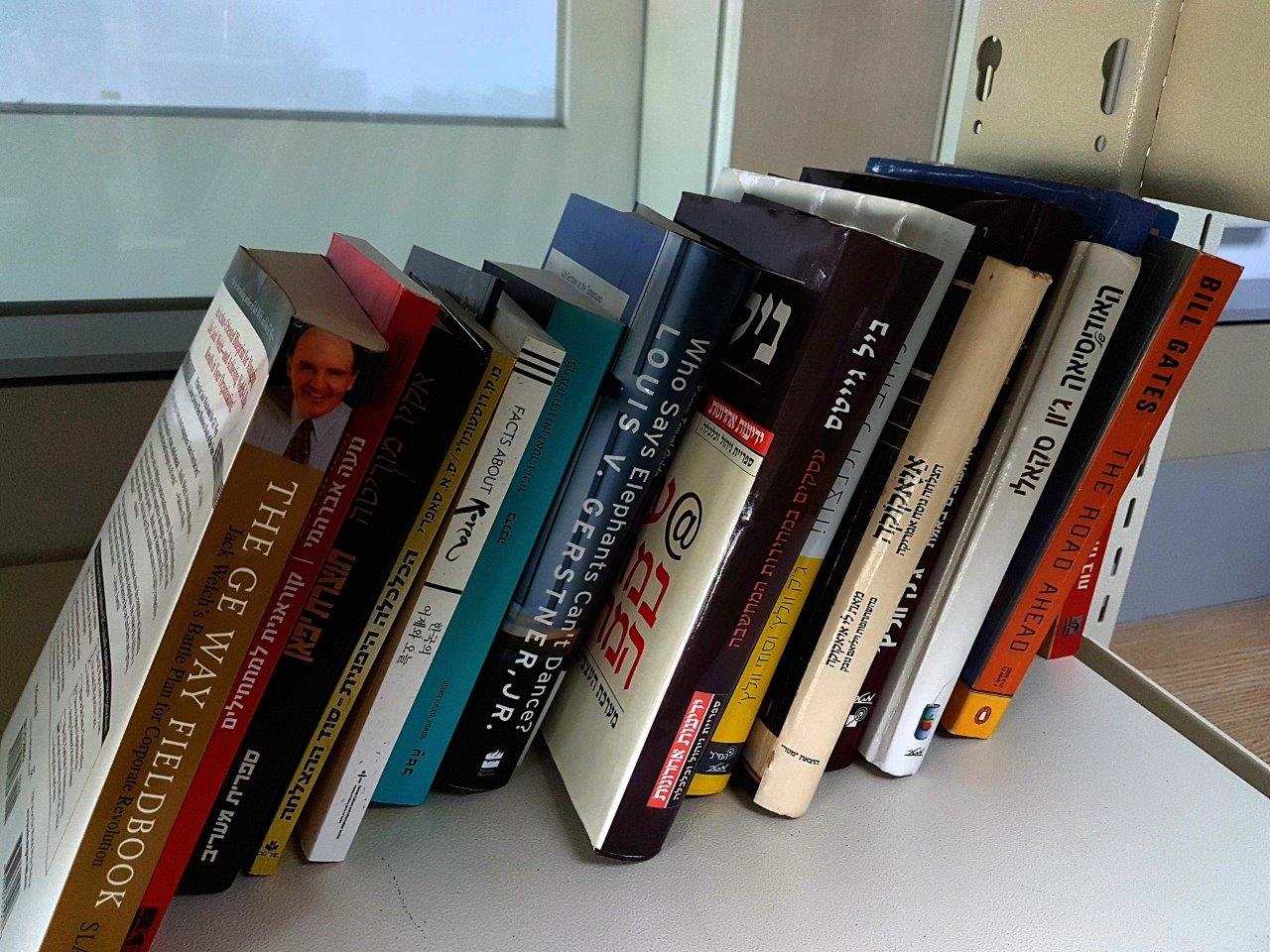} &
        \includegraphics[width = 0.24\textwidth]{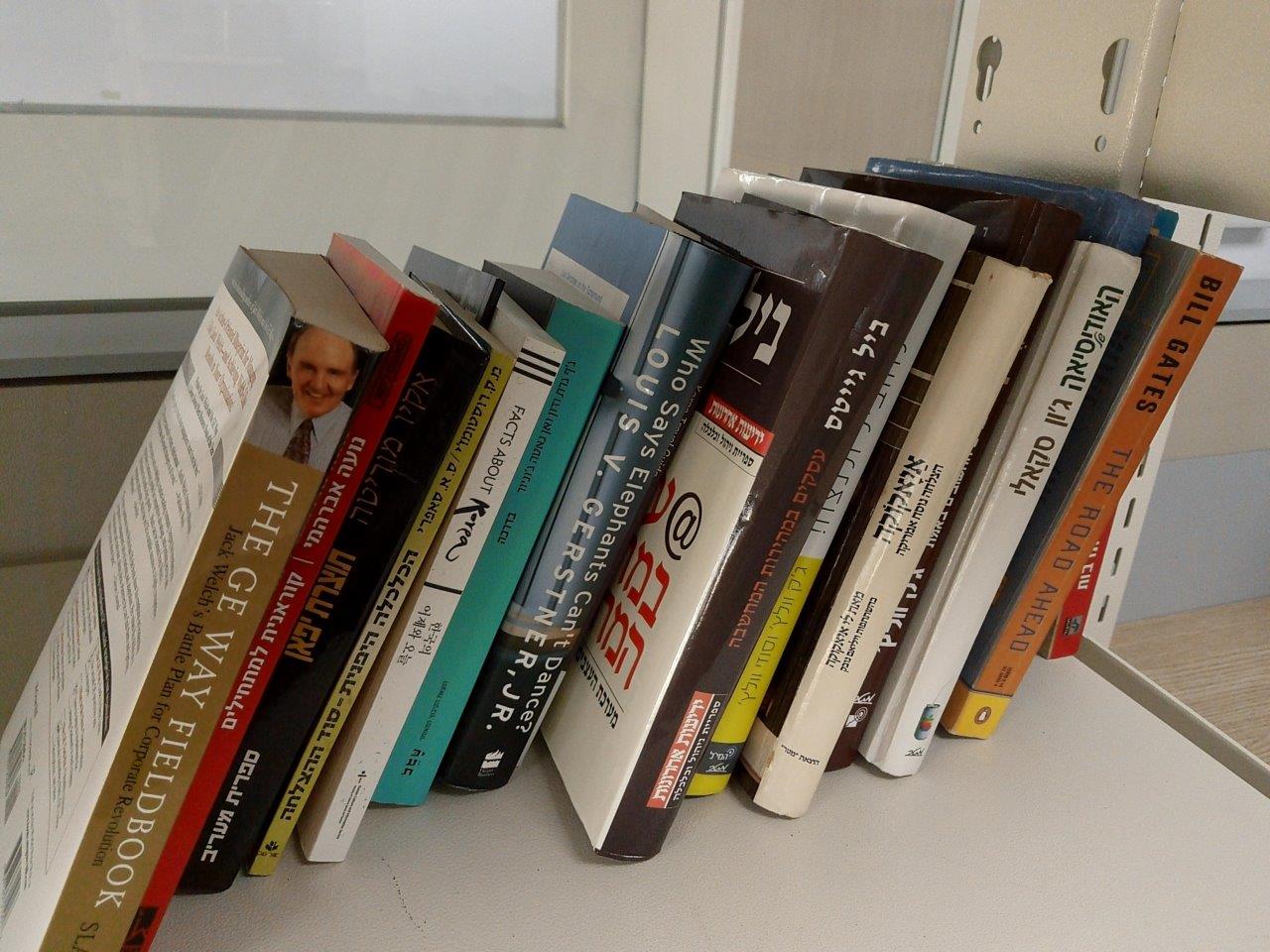} \\       
    \end{tabular}         
    \begin{tabular}{@{\hskip 0.005\textwidth}l@{\hskip 0.01\textwidth}l@{\hskip 0.01\textwidth}l@{\hskip 0.01\textwidth}l@{\hskip 0.01\textwidth}l@{\hskip 0.01\textwidth}l@{\hskip 0.01\textwidth}l@{\hskip 0.01\textwidth}c}   
          \includegraphics[width = 0.115\textwidth]{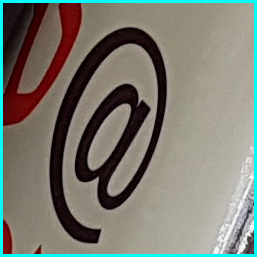} &
          \includegraphics[width = 0.115\textwidth]{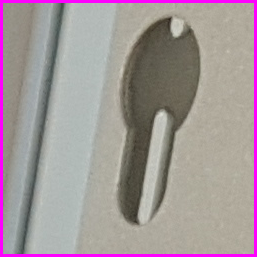} &
          \includegraphics[width = 0.115\textwidth]{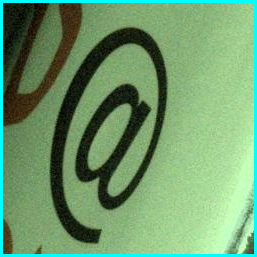} &
          \includegraphics[width = 0.115\textwidth]{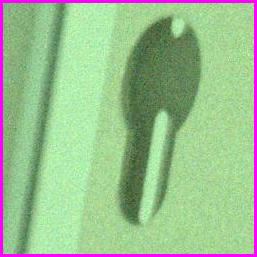} &
          \includegraphics[width = 0.115\textwidth]{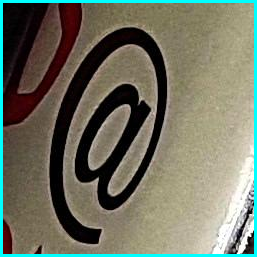} &
          \includegraphics[width = 0.115\textwidth]{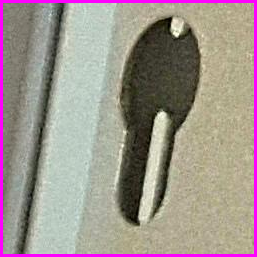} &
          \includegraphics[width = 0.115\textwidth]{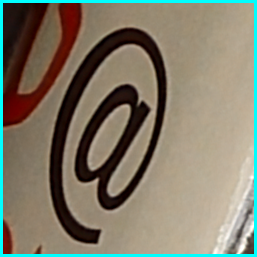} &
          \includegraphics[width = 0.115\textwidth]{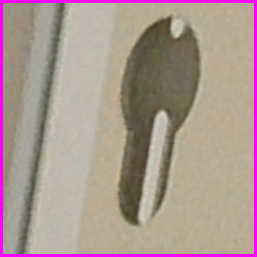} \\    
    \end{tabular}           
    \begin{tabular}{@{\hskip 0.005\textwidth}c@{\hskip 0.07\textwidth}c@{\hskip 0.07\textwidth}c@{\hskip 0.07\textwidth}c}
        {Well-lit (ground truth)}&
        {\ Low-light raw input\ \ }&
        {\ \ \ \ \ Samsung S7 \ \ \ \ \ }& 
        {\ \ \ \ \ \ \ \ \ DeepISP \ \ \ \ \ \ \ }\\ 
    \end{tabular}      
    \smallskip  
    \caption{\small \textbf{More examples of the end-to-end low-light image processing.} From left to right: a ground truth well-lit image, raw input low-light image (for visualization purposes, after demosaicing by bilinear interpolation), output of the Samsung S7 ISP (after histogram stretch), and of the proposed DeepISP.}
    \label{fig_teaser_extra}
\end{figure*}

To evaluate the reconstruction results we use mean opinion score (MOS), which has been generated using the Amazon Mechanical Turk for both full images and patch level as specified above. 
A total of $200$ ratings have been collected for each image ($200$ per version of an image, i.e., DeepISP output, Samsung S7 output and the well-lit ground truth): $100$ ratings for $10$ random patches and additional $100$ for the full image. 
Fig.~\ref{fig_isp_dataset_results_bar_graph} presents the evaluation results. 
For the patch level, DeepISP MOS is $2.86$ compared to Samsung S7 ISP which has $2.71$ on the same images. 
The DeepISP MOS for full images is $4.02$ compared to $3.74$ achieved by Samsung S7 ISP. The former result is only slightly inferior to the MOS $4.05$ that is given to the well-lit images. 
It is also evident that the visual quality score predicted by DeepIQA \cite{bosse2016deep} corresponds well to the human evaluation with scores of $3.72$, $3.92$ and $4.02$ for the Samsung S7 ISP, DeepISP and the well-lit scene, respectively. Figures \ref{fig_teaser} and \ref{fig_teaser_extra} present a selection of visual results. \reb{Since Samsung's low-light images are quite dark (to suppress visible noise), we present (and evaluated) all images after a simple histogram stretching to have a fair comparison. The luminance channel histogram has been stretched to cover the range $\left[0-255\right]$ with 5\% saturation at the high and low boundaries of the range. Samsung's original low-light images (after the Samsung ISP but before our histogram) are rated about 1 point lower on the MOS compared to the same images after histogram stretching.}

\textbf{DeepISP for well-lit images.}
Despite the fact that our work is focused on ISP for low-light images, we also show that our proposed model can be used for well-lit images. 
For this purpose, we trained a similar model for well-lit images.  
However, unlike the low-light case, where we had a higher quality image that we could use as the ground truth, in this case, we have only the raw version and its processed JPEG version from the Samsung ISP. With these limitations, we trained our network to mimic the Samsung ISP, having the ``well-lit'' raw image (captured in fully automatic mode) as the input to the network and the JPEG as its target ``ground truth''.

\begin{figure}[tb] 
\begin{center}
   \includegraphics[width=0.9\linewidth]{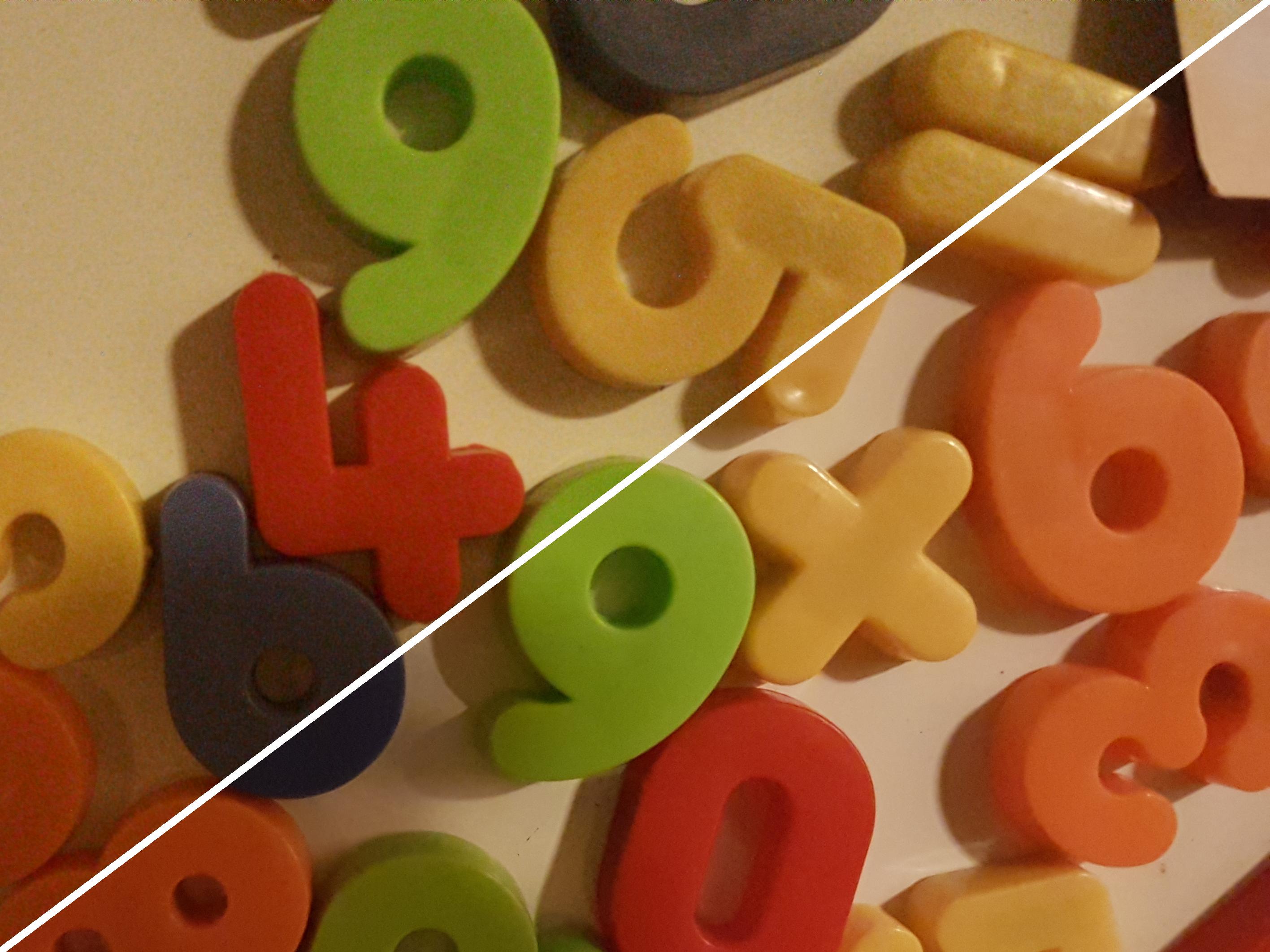} \\
   \smallskip
   \includegraphics[width=0.9\linewidth]{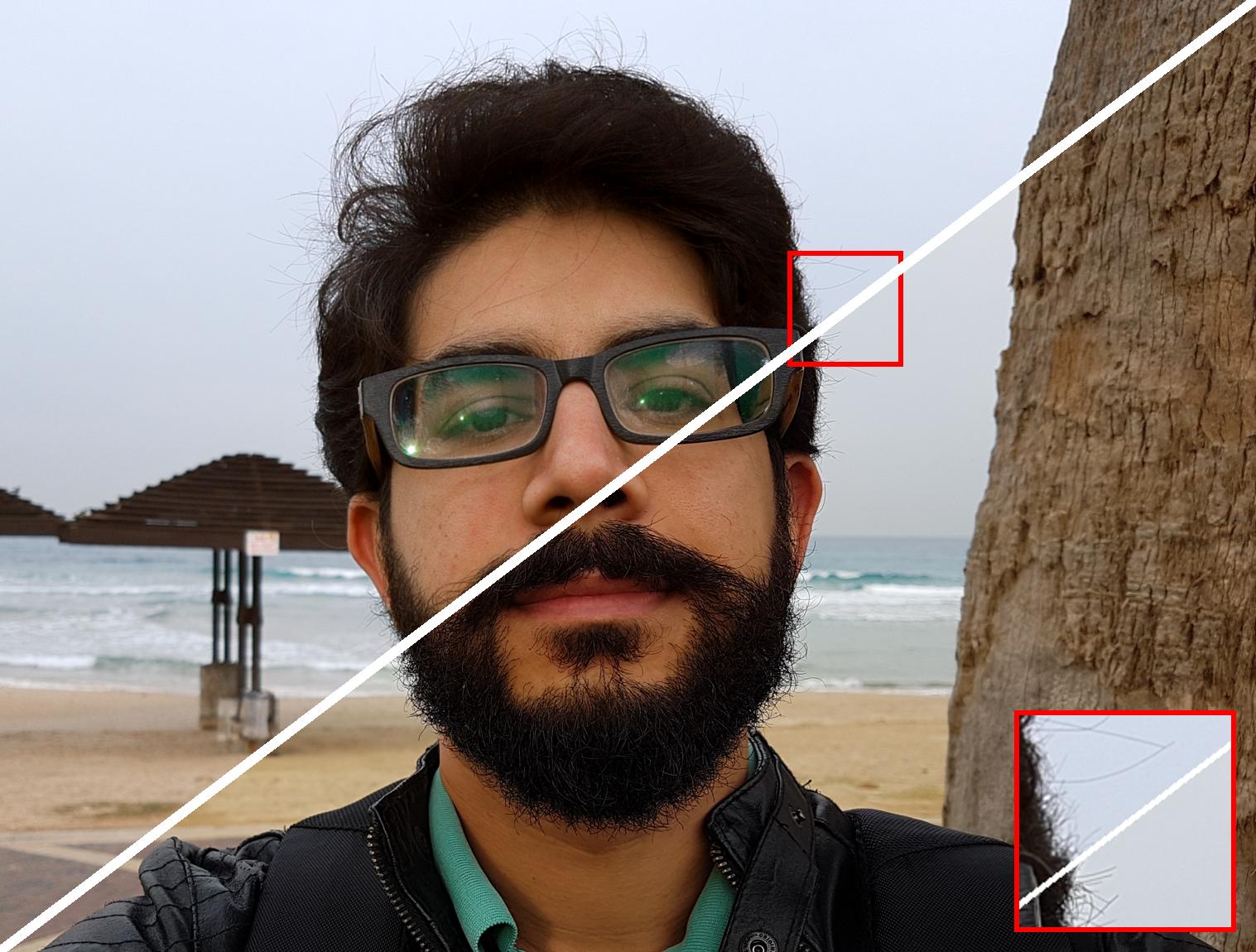}
\end{center}
       \caption{\textbf{Mimicking an ISP.} \reb{A model trained to mimic Samsung's ISP. Both RAW input image and JPEG (ground truth) are captured in fully automatic mode. Upper-left triangle is the output of DeepISP. Lower-right is Samsung's output.}}
    \label{fig_mimic_isp}
\end{figure}

For training the network we use the same training procedure, same hyper-parameters and number of epochs as for the low-light processing experiment described above. The initial transformation for the high-level part--$W_i$, was computed for these inputs in the same way described for the low-light case. 
Note that we trained the network in sub-optimal conditions as we had only the JPEG images as the target output and not (noiseless) higher quality image. 
Nevertheless, the model was able to mimic the ISP and generate pleasant looking images which are indistinguishable from the ground truth when examining the full-scale image and are close to the ground truth when examining details (Fig.~\ref{fig_mimic_isp}). One should keep in mind that the ground-truth images used in the training are the upper bound for what we can expect to achieve.

This experiment demonstrates that a neural model can learn to mimic an ISP given as a black box. Moreover, the good results achieved in this setting combined with the good low-light processing results achieved when the high-quality ground truth was given, we argue that our DeepISP architecture is likely to produce a better output when given a higher-quality ground truth at training.





\begin{figure}[tb] 
\begin{center}
   \includegraphics[width=0.9\linewidth]{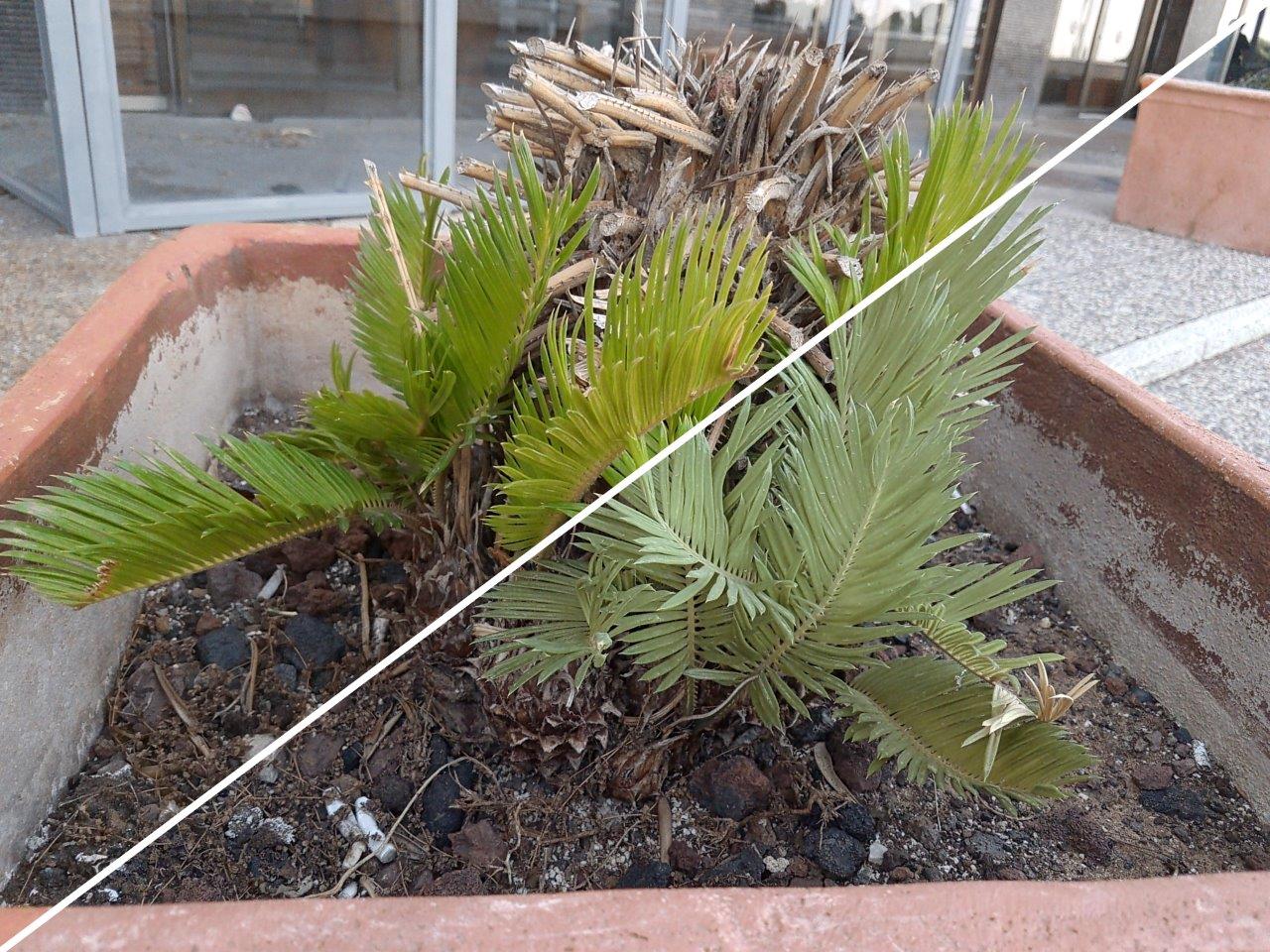} \\
   \smallskip
   \includegraphics[width=0.9\linewidth]{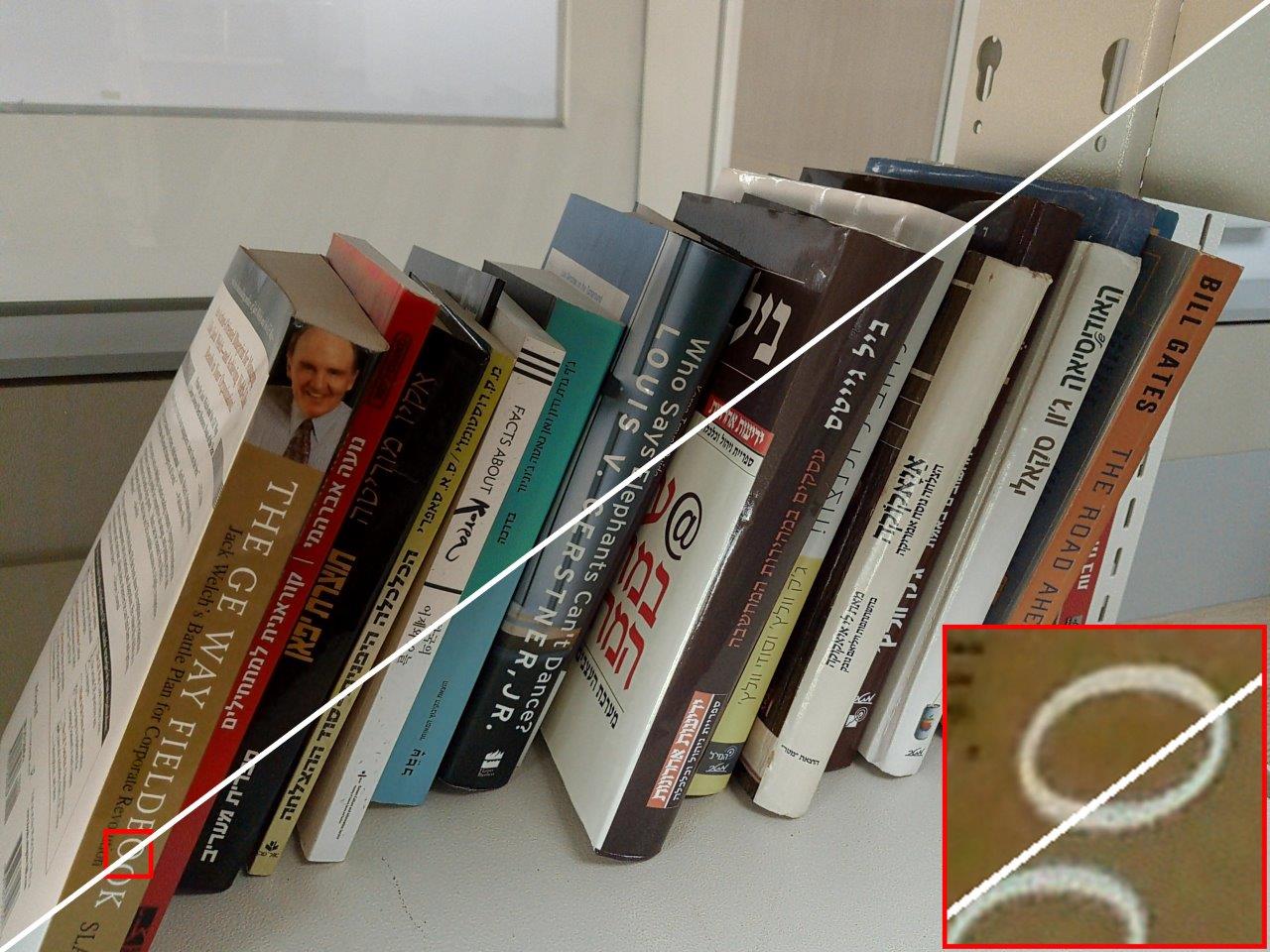}
\end{center}
       \caption{\reb{\textbf{Importance of shared features.} Upper-left triangle: output of DeepISP. Lower-right triangle: a modified DeepISP with a severed connection between the low-level and high-level stages. When features are not shared, the model often fails to generate good-looking colors.}}
    \label{fig_cutting_ll_and_hl}
\end{figure}

\textbf{Importance of shared features.}
We trained a modified DeepISP to study the effect of simultaneous learning of low- and high-level corrections. 
\reb{
In this experiment we show that given the same budget (number of layers and number of parameters) we get inferior results when information is not shared.
}
When the connection between the low-level and high-level stages was severed (no shared features, i.e., the high-level part just gets the output image of the low-level stage), we observed degraded image quality (Fig.~\ref{fig_cutting_ll_and_hl}).

\section{Conclusion}

\minorrev{
In this paper, we presented DeepISP -- a novel end-to-end deep learning model that brings us one step forward towards having a learned network that may replace the full ISP of a digital camera. Although deep learning has been applied previously to different tasks in the image processing pipeline, to the best of our knowledge, this is the first attempt to tackle all of them simultaneously. Such an approach has the advantage of sharing information while performing different tasks. This has the potential to lower computational costs compared to the case when each processing step is performed independently. The steps that are excluded in the current network and require further exploration are removing camera shake/blur, adding options for HDR and adapting the network for various levels of noise. We believe that by adding them, which should be done as a follow-up work, one may be able to replace the current ISP in modern cameras with a learned one.
}


The DeepISP model demonstrated its ability to generate visually compelling images from raw low-light images.
The output of the Samsung S7 ISP was used as the reference both with low-light and well-lit raw inputs. To evaluate the image quality, we relied on both human ratings (MOS) using Amazon's Mechanical Turk and DeepIQA \cite{bosse2016deep} -- a DL model that predicts subjective image quality. In the human evaluation of full images, DeepISP images scored $7\%$ higher than the manufacturer ISP and just $0.7\%$ below the equivalent well-lit images. Similar trends were observed with the DeepIQA measure.


We also tested the performance of our solution on a low-level image processing task to measure its performance in terms of an objective metric (PSNR). We considered the problem of joint denoising and demosaicing. Our technique outperformed the state-of-the-art (both DL-based and axiomatic methods) by $0.72dB$ PSNR on the Panasonic MSR Dataset. For the task of training on the Panasonic MSR Dataset and testing on the Canon MSR Dataset, our network demonstrated its ability to generalize well, outperforming by $1.28dB$ PSNR over the previous solutions.

While there is no ``agreed objective metric'' for measuring the general quality of an image in the end-to-end case, our study suggests that networks for measuring quality such as DeepIQA can serve as such. As observed in our experiments, DeepIQA score is well-correlated with human perception. Future work may use a network similar to DeepIQA to improve the perceptual quality of our network even further. It can be used as part of the loss and it gradients can be propagated through the networks. This may serve as an alternative to the popular adversarial loss, which is used in other studies.

\changes{Another future research direction is optimizing the ISP not just for human perception but also for improving performance of higher level algorithms using these images as input, such as object classification. In \cite{buckler2017reconfiguring} it has been shown that skipping different ISP tasks affect the performance of classification models (trained for these images). That suggests that the ISP tasks are important for the performance of higher level algorithms and thus might be optimized for them.
}

\section*{Acknowledgment}

This research was partially supported by ERC-StG SPADE PI Giryes and ERC-StG RAPID PI Bronstein. The authors are grateful to NVIDIA's hardware grant for the donation of the Titan X that was used in this research.

\bibliographystyle{IEEEtran}
\bibliography{./bibtex/bib/IEEEabrv,./bibtex/bib/deepisp}

\begin{IEEEbiography}[{\includegraphics[width=1in,height=1.25in,clip,keepaspectratio]{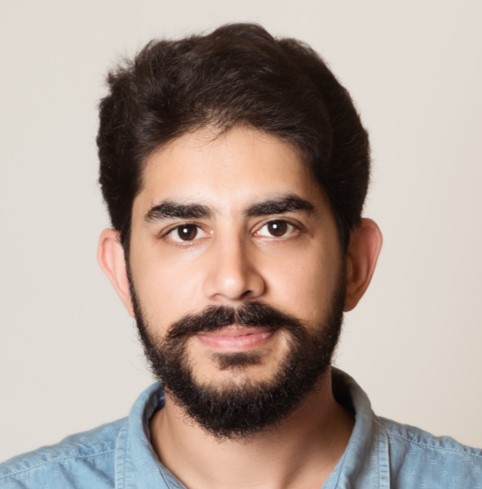}}]{Eli Schwartz}
holds a M.Sc in Electrical Engineering from Tel Aviv University, Israel (2018) and B.Sc in Electrical Engineering from the Technion - Israel Institute of Technology (2013). His Master's thesis deals with application of deep learning models for the full image processing pipeline.

Eli has authored 4 academic papers with publications in top journals and conferences and 5 patents/patent applications.

He is currently with IBM Research AI conducting publishable research on deep learning for object recognition and detection (2017-present).

Eli has co-founded and served as the CTO of a robotics startup - Inka Robotics that developed a world first tattooing robot (2015-2017). Previously, Eli was with Microsoft, developing algorithms for the HoloLens AR headset (2013-2016). Eli also has experience in the chip design industry, having worked for Qualcomm (2011-2013) and IBM (2008-2011).
\end{IEEEbiography}

\begin{IEEEbiography}[{\includegraphics[width=1in,height=1.25in,clip,keepaspectratio]{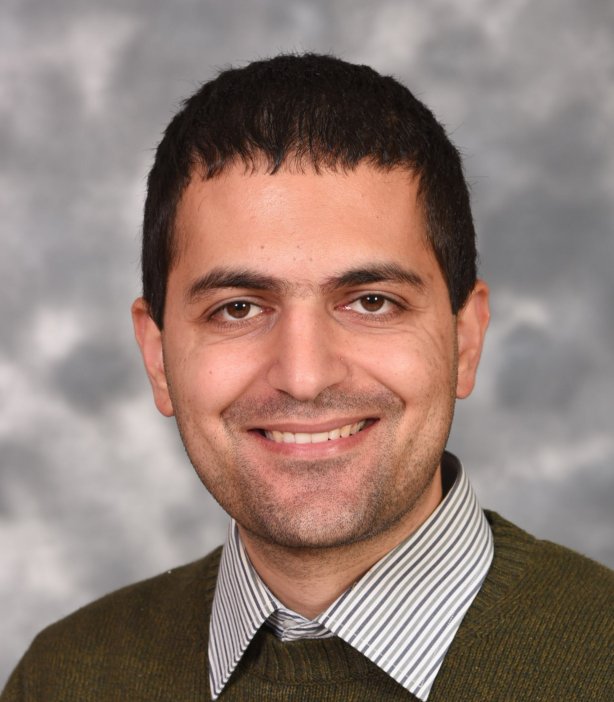}}]{Raja Giryes}
is a senior lecturer in the school of electrical engineering at Tel Aviv University. He received the B.Sc (2007), M.Sc. (supervision by Prof. M. Elad and Prof. Y. C. Eldar, 2009), and PhD (supervision by Prof. M. Elad 2014) degrees from the Department of Computer Science, The Technion - Israel Institute of Technology, Haifa. Raja was a postdoc at the computer science department at the Technion (Nov. 2013 till July 2014) and at the lab of Prof. G. Sapiro at Duke University, Durham, USA (July 2014 and Aug. 2015). His research interests lie at the intersection between signal and image processing and machine learning, and in particular, in deep learning, inverse problems, sparse representations, and signal and image modeling.

Raja received the ERC-StG grant, Maof prize for excellent young faculty (2016-2019),
JVCI best paper award honorable mention, VATAT scholarship for excellent postdoctoral fellows (2014-2015), Intel Research and Excellence Award (2005, 2013), the Excellence in Signal Processing Award (ESPA) from Texas Instruments (2008) and was part of the Azrieli Fellows program (2010-2013). He has co-organized workshops and tutorials on deep learning in leading conference such as ICML 2016 and 2018, ICCV 2015 and 2017, CVPR 2016 and 2017, EUSIPCO 2016, ACCV 2016 and CDC 2017.
\end{IEEEbiography}

\begin{IEEEbiography}[{\includegraphics[width=1in,height=1.25in,clip,keepaspectratio]{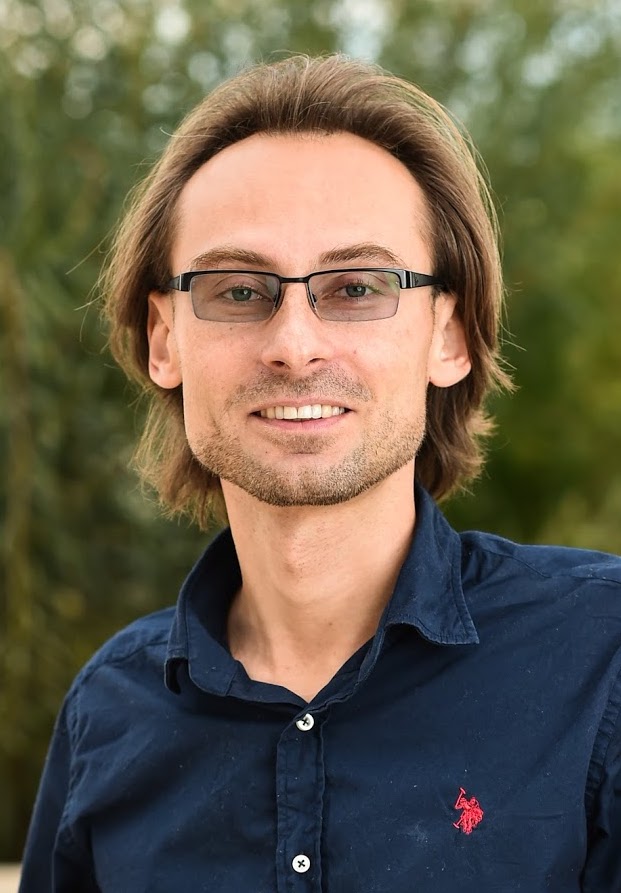}}]{Alex Bronstein}
is an associate professor of computer science at the Technion – Israel Institute of Technology a principal engineer at Intel Corporation. His research interests include numerical geometry, computer vision, and machine learning. Prof. Bronstein has authored over 100 publications in leading journals and conferences, over 30 patents and patent applications, the research monograph "Numerical geometry of non-rigid shapes", and edited several books. Highlights of his research were featured in CNN, SIAM News, Wired. Prof. Bronstein is a Fellow of the IEEE for his contribution to 3D imaging and geometry processing. In addition to his academic activity, he co-founded and served as Vice President of technology in the Silicon Valley start-up company Novafora (2005-2009), and was a co-founder and one of the main inventors and developers of the 3D sensing technology in the Israeli startup Invision, subsequently acquired by Intel in 2012. Prof. Bronstein's technology is now the core of the Intel RealSense 3D camera integrated into a variety of consumer electronic products. He is also a co-founder of the Israeli video search startup Videocites where he serves as Chief Scientist.
\end{IEEEbiography}

\end{document}